\documentclass[aip,jcp,preprint]{revtex4-1}

\usepackage{amsmath}
\usepackage{amssymb}
\usepackage{graphicx}

\usepackage{dcolumn}
\usepackage{bm}
\usepackage{longtable}
\usepackage[usenames, dvipsnames]{color}
\usepackage{subfigure}
\usepackage{multirow}
\usepackage{dsfont}
\usepackage{ulem}

\DeclareMathOperator*{\argmin}{argmin}

\begin{document}

\preprint{AIP/DUCHEMIN}

\title{Separable Resolution-of-the-Identity with All-Electron Gaussian Bases: Application to Cubic-scaling RPA}

\author{Ivan Duchemin}
\email{ivan.duchemin@cea.fr}
\affiliation{Univ. Grenoble Alpes, CEA, INAC-MEM-L\_Sim, 38000 Grenoble, France}
\author{Xavier Blase}
\affiliation{Univ. Grenoble Alpes, CNRS, Inst NEEL, F-38042 Grenoble, France}

\date{\today}

\begin{abstract}

We explore a separable resolution-of-the-identity formalism built on quadratures over limited sets of real-space points designed for all-electron calculations. Our implementation preserves in particular the use of common atomic orbitals and their related auxiliary basis sets. The set up of the present density fitting scheme, i.e. the calculation of the system specific quadrature weights, scales cubically with respect to the system size. Extensive accuracy tests are presented for the Fock exchange and MP2 correlation energies. We finally demonstrate random phase approximation (RPA) correlation energy calculations with a scaling that is cubic in terms of operations, quadratic in memory, with a small crossover with respect to our standard RI-RPA implementation.

\end{abstract}

\maketitle

\section{Introduction}

The resolution-of-identity (RI)\cite{Whi73,Dun79,Min82,Als88,Fey93,Vah93,Klo02}  stands as a central technique in quantum chemistry, relying on the expansion of $\phi_n  \phi_m  $ co-densities over an auxiliary atomic basis set $\lbrace \beta \rbrace$ that scales linearly with the number of atoms.
Even though the auxiliary basis sets are typically three times larger than the corresponding atomic orbital (AO) basis sets supporting the $ \phi_n $ Hartree-Fock or Kohn-Sham molecular orbitals, this represents a considerable saving both in terms of memory and number of operations to be performed, which comes at the price of a moderate accuracy loss.

While RI was introduced initially to facilitate the calculation of 2-electron 4-center $(mn|kl)$ Coulomb integrals, operators that finds an exact expression in the product space between occupied and virtual eigenstates can also be expressed more compactly using auxiliary bases. As such, the independent-electron density-density susceptibility $\chi_0({\bf r},{\bf r}';\omega)$ is of central importance to the present study. In particular, the scaling of the related random phase approximation (RPA) approach, \cite{Boh53,Gun76,Lan77} a popular low-order perturbative approach to correlation energies beyond density functional approximations, can be reduced from $\mathcal{O}(N^6)$ to $\mathcal{O}(N^4)$. \cite{Esh10}

The computational efficiency and accuracy of RI techniques strongly depend on the scheme adopted to build the appropriate coefficients expressing the $ \phi_n  \phi_m $ co-densities on the auxiliary basis. The original density-fitting (RI-SVS) approach\cite{Dun79,Als88,Fey93} expresses these coefficients as a direct overlap $\langle \phi_m \phi_n | \beta \rangle$, requiring the calculation of the sparse $\langle \alpha {\alpha}' | \beta \rangle$ coefficients, with $\lbrace \alpha \rbrace$ being the AO basis set used to expand the molecular orbitals  $ \phi_n $. The now widely adopted Coulomb-fitting  (RI-V) approximation \cite{Min82,Vah93} requires on the other hand the calculation of the denser 3-center Coulomb integrals $(\alpha {\alpha}' | \beta )$, displaying much less sparsity than $\langle \alpha {\alpha}' | \beta \rangle$ integrals due to the long-range nature of the Coulomb operator. The Coulomb-fitting formalism is known to be more accurate than the density-fitting scheme for auxiliary basis sets of similar sizes, \cite{Ren12,Duchemin16}  bringing to the standard issue of the trade-off between accuracy and computational/memory costs. The use of short-range or attenuated Coulomb operators, \cite{Jun05,Rei08,Sod08,Ihr15} or corrective techniques such as ``multipole-preserving" constraints to the density-fitting scheme,\cite{Als88,Duchemin16} allows one to tune the accuracy-to-cost ratio between these two standard RI approximations.

In conjunction with other powerful techniques, such as the Laplace transform, \cite{Alm91}   exploiting the sparsity of the RI-SVS density-fitting $\langle \alpha {\alpha}' | \beta \rangle$ coefficients in the limit of large systems was shown to allow cubic-scaling RPA calculations. \cite{Wil16}  As a trade-off between accuracy and efficiency, a Coulomb-attenuated variation of the Coulomb-fitting  (RI-V) RPA was recently explored to obtain a low-scaling formulation, \cite{Lue17}  exploiting further the decay properties in real-space of the Laplace-transformed "pseudo" density matrices  expressed in the AO basis. \cite{Has93,Aya99,Kal15,Sch16}
The efficiency of this latter family of approaches depends on the electronic properties of the system of interest, and are different in nature from the sparsity associated with specific resolution-of-identity formalisms. 

In the present study, we explore an alternative approach for reducing computational and memory loads  by assessing on a large set of molecules the merits of a separable resolution-of-the-identity formalism relying on a density fitting scheme over compact sets of real-space points $\lbrace {\bf r}_k \rbrace$. Our approach preserves the use of standard Gaussian atomic orbitals and auxiliary basis sets for all-electron calculations, targeting the accuracy of the Coulomb-fitting (RI-V) formalism.  The set-up of the fitting procedure scales cubically with system size.  The accuracy of our approach is first validated by an extensive benchmark of the exchange and MP2 correlation energies over a large set of molecules. Combined with the Laplace transform technique, and following the so-called space-time approach for calculating the susceptibility operator,\cite{Roj95,Kal14} the calculation of the RPA correlation energy within the present real-space quadrature approach is shown to scale cubically in terms of operations and  quadratically in terms of memory, without invoking any sparsity or localization properties. The accuracy of the present real-space RI-RPA formalism is further shown  to  match that of the standard quartic-scaling Coulomb-fitting RI-RPA calculations for a large set of molecules including the oligoacenes, $C_{60}$ and a larger octapeptide angiotensin II molecule (146 atoms including 71 H atoms) proposed by Eshuis and coworkers in the early days of RI-RPA implementations. \cite{Esh10}  

\section{Theory}

In this Section, we briefly outline the standard RI-V and RI-SVS approximations, introducing the notations used throughout the paper. We then discuss separable resolution-of-identities and present our specific implementation preserving the use of standard atomic and auxiliary basis sets.
The present approach relies on weighted real-space $\delta({\bf r}-{\bf r}_k)$-functions to express the density fitting coefficients, relating co-densities to auxiliary basis functions through real-space quadratures. 
The scheme to optimize the distribution of ${\bf r}_k$ and related weights is presented, and  compared to other real-space quadrature formalisms. We demonstrate in particular that the computational cost associated with the setup of the present RI approach scales cubically with the system size. 
We show then how such a separable RI allows to obtain cubic-scaling RPA with low crossover with respect to standard quartic RI-RPA formalism when combined with the Laplace transform technique.
We conclude this Section by presenting the technical details and parameters adopted in this study to perform the calculations illustrating the accuracy and scaling properties of the present approach.

\subsection{Standard Resolution of the Identity}

The resolution-of-identity (RI) approximation\cite{Whi73,Dun79,Min82,Als88,Fey93,Vah93,Klo02} relies on the expansion of molecular orbitals co-densities $ \phi  \phi' $ over an auxiliary basis set $\lbrace \beta \rbrace$, namely: 
\begin{equation}
\begin{split}
\phi(\mathbf{r})\phi'(\mathbf{r}) \simeq &\sum_{\beta} \mathcal{F}_{\beta}(\phi\phi') \; \beta(\mathbf{r}) \\
\doteqdot &  \, \mathcal{F}(\phi\phi';\mathbf{r}) \\
\label{generic}
\end{split}
\end{equation}The fit $\mathcal{F}$ is realized through an ensemble of measures $\{\mathcal{F}_{\beta}\}$,  mapping the $ \phi \phi' $ product-space to the $ \beta $ auxiliary subspace defined so as to scale linearly with the number of atoms. Typical examples of such procedures are the standard RI-V and RI-SVS fitting approaches that use respectively:   
\begin{eqnarray}
\mathcal{F}_{\beta}^{V}(\phi\phi')  = \sum_{\beta'} [V^{-1}]_{\beta\beta'} \; (\beta'|\phi\phi') \\
\mathcal{F}_{\beta}^{SVS}(\phi\phi') = \sum_{\beta'} [S^{-1}]_{\beta\beta'} \; \langle\beta'|\phi\phi'\rangle 
\end{eqnarray}
where $V$ and $S$ represent respectively the Coulomb $(\beta|\beta')$ and overlap $\langle\beta|\beta'\rangle$ matrices associated with the  auxiliary basis set, and $[X^{-1}]_{\beta\beta'}$ denotes the $(\beta,\beta')$ entry of the $X$ inverse matrix. To explicitly define our $\langle \cdot | \cdot \rangle$ and $( \cdot | \cdot )$ notations, we write:
\begin{eqnarray*}
(\beta|\phi\phi') &=& \iint d{\bf r}d{\bf r}' \; \frac{ \beta({\bf r}) \phi({\bf r}')\phi'({\bf r}')  }{| {\bf r} - {\bf r}' | } \\
\langle\beta |\phi\phi'\rangle  &=& \int d{\bf r} \;\beta({\bf r}) \phi({\bf r})\phi'({\bf r}) 
\end{eqnarray*}
As shown in Ref.~\citenum{Duchemin16}, both fitting techniques can be combined, preserving the Coulomb-fitting RI-V approach for low angular momentum auxiliary $ \beta $ atomic orbitals. As emphasized here above, the number of $\langle\beta|\alpha\alpha'\rangle$ overlap matrix elements in the RI-SVS approximation scales linearly with system size, offering a first strategy for reducing computational cost and memory thanks to sparsity. On the contrary, the number of $(\beta |\alpha\alpha')$ Coulomb integrals scales quadratically so that sparsity, or sparse tensor algebra,  is difficult to exploit within the more accurate Coulomb-fitting (RI-V) approach.

\subsection{Separable RI}

A separable expression for the resolution-of-the-identity can be obtained through a set of separable measures $\{\langle f_k|\}$ on the co-densities, namely:
\begin{equation}
\begin{split}
\mathcal{F}_{\beta}(\phi\phi')  & = \sum_{k} [M]_{\beta k} \; \langle f_k|\phi\phi'\rangle\\
& = \sum_{k} [M]_{\beta k} \; \langle f_k|\phi\rangle \; \langle f_k|\phi'\rangle\\
\end{split}
\end{equation}
where the coefficients of $M$ have yet to be defined. 
Though it is not the only option, a trivial way to obtain separable measures is to work with $\delta(\mathbf{r}-\mathbf{r}_k)$ distributions centered on  a set of $N_{k}$ real-space locations $\{\mathbf{r}_k\}$. Working with real-space (RS) measures, the $\mathcal{F}^{RS}$ density fitting procedure takes then the simple form  
\begin{equation}
\begin{split}
\mathcal{F}^{RS}_{\beta}(\phi\phi') &  = \sum_{k} [M]_{\beta k} \langle \delta(\mathbf{r}-\mathbf{r}_k) |\phi\phi'\rangle\\
& = \sum_{k} [M]_{\beta k} \; \phi(\mathbf{r}_k) \; \phi'(\mathbf{r}_k) 
\label{rirsdef}
\end{split}
\end{equation}  
The clear advantage of the separability is that the two molecular orbitals $\phi \phi'$, originally entangled in e.g. the $\mathcal{F}_{\beta}^{V}(\phi\phi')$ RI-V fitting coefficients through the $(\beta|\phi \phi')$ Coulomb integrals, are now disentangled. This will prove crucial in the calculation of linear-response or perturbation theory related quantities where summations over occupied/virtual pairs have to be performed as shown below in the case of the calculation of the RPA correlation energy. We emphasize however that, while relying on discrete values of the molecular orbitals in real-space, the present approach remains a resolution-of-the-identity in the sens that  physical continuous quantities such as the co-densities, linear-response operators   (e.g. suceptibility), etc. are defined everywhere in space in terms of the $ \beta $ auxiliary basis functions. 

Amongst existing formalisms adopting real-space quadrature strategies, several studies focused directly on 2-electron Coulomb integrals. The chain-of-sphere (COSX) semi-numerical approach to exchange integrals,\cite{Neese09} building on Friesner’s pioneering pseudo-spectral approach,\cite{Fri91} develops only one of the two co-densities forming 2-electron Coulomb integrals over a real-space grid. Alternatively, the  Least Square Tensor Hypercontraction (LS-THC) formalism \cite{Parrish12} fully develops the 2-electron integrals as a quadrature over real-space grid points, with an $\mathcal{O}(N^4)$ computational complexity associated with the  establishment of the quadrature. 

On the other hand, the Interpolative Separable Density Fitting (ISDF) from Lu and coworkers provides a $\mathcal{O}(N^3\log(N))$ separable fit tensor by using Fourier transform and random projection techniques to select the $\mathbf{r}_k$ points and define the corresponding auxiliary densities, with a proof-of-concept application to a simple model system.\cite{Lu14,Ying16}  The present work adopts as well a separable form for the fit tensor (Eq.~\ref{rirsdef}) through separable measures along real-space positions, but differs in the way the real-space locations $\mathbf{r}_k$ and their associated weights $[M]_{\beta k}$ are constructed, leading to a $\mathcal{O}(N^3)$ quadrature determination process. In addition, we preserve the auxiliary atomic basis sets in use in quantum chemistry. The accuracy and efficiency of our approach is further  benchmarked on a large number of molecular systems.

\subsection{Quadrature weight determination}

Assuming that the $\mathbf{r}_k$ locations are known (see below), the determination of the $[M]_{\beta k}$ coefficients in  Eq.~\ref{rirsdef} are obtained from the following minimization condition:
\begin{equation}
\argmin_{M} \sum_{\rho,\beta}\Big( \mathcal{F}_\beta^{RS} (\rho)  - \mathcal{F}_\beta^{V} (\rho) \Big)^2  \label{eq:fit_LSQR}
\end{equation}Namely, we aim to equate the fitting functions of the present RI-RS approach with that of the Coulomb fitting RI-V scheme.  The accuracy of the present RI-RS approach is thus targeting that of the RI-V approach. The set of test co-densities $\lbrace \rho \rbrace$ typically spans the $\lbrace \alpha \rbrace \otimes \lbrace \alpha \rbrace$ product-space, even though it can be adjusted depending on the problem being addressed. Interestingly,  the present fitting scheme allows to recover at reduced cost the otherwise $\mathcal{O}(N^4)$ LS-THC factorization (see Supporting Information SI \cite{suppinfo}). The solution of Eq.~\ref{eq:fit_LSQR} can indeed  be achieved with an advantageous $\mathcal{O}(N^3)$ computational complexity, as demonstrated now.

In order to detail our fitting procedure, we adopt the following matrix notations: $[D]_{k\rho}=\rho(r_k)$, $[F]_{\beta\rho}=\mathcal{F}_\beta^{V} (\rho)$. 
Due to the localization properties of the atomic orbitals, the number of atomic orbital products scales linearly with system size, and thus the number $N_\rho$ of test codensities in the test set can be considered $\propto N_\alpha$. As a result, the matrices D ($N_k\times N_\rho$), F ($N_\beta\times N_\rho$) as well as the matrix M ($N_\beta\times N_k$) of Eq.~\ref{rirsdef} are all $\mathcal{O}(N^2)$ tensors. The fit equation Eq.~\ref{eq:fit_LSQR} can then be formulated as:
\begin{equation}
\argmin_{M} \Big|\Big| M\cdot D  - F \Big|\Big|   \label{eq:fit_LSQR_2}
\end{equation}which leads to the standard least-square estimator:
\begin{equation}
M = F\cdot D^\dag \cdot ( D\cdot D^\dag )^{-1}\label{eq:fit_LSQR_3}
\end{equation} involving only matrix multiplications and inversions. Computation of $ ( D\cdot D^\dag )^{-1}$ could prove problematic if done explicitly. The term $ D\cdot D^\dag$ is positive, but has not guarantee to be definite. On the other hand, due to the large number $N_\rho$ of test co-densities, application of the standard SVD technique to extract the pseudo-inverse based estimator leads to rather significant prefactors to the otherwise  $\mathcal{O}(N^3)$ pseudo-inverse procedure. We take a side approach by combining simple balancing and Tikhonov $L_2$ regularization. \cite{Tikho95}  We first balance the problem by normalizing the rows of $D$, writing $\widetilde{D}=d\cdot D$ where $d$ is a diagonal matrix and the diagonal terms $[\widetilde{D}\cdot \widetilde{D}^\dag]_{kk}=1$. The pseudo inverse is then calculated as:
\[
 ( D\cdot D^\dag )^{-1} \simeq d\cdot ( \widetilde{D}\cdot \widetilde{D}^\dag+\epsilon\mathds{I})^{-1}\cdot d
\]where the $L_2$ regularization parameter $\epsilon$ is adjusted to a small value to maintain definiteness of the problem and ensure numerical stability of the inverse. We identified the value $\epsilon=4 \times 10^{-7}$ as a reasonable parameter for double precision arithmetic and kept this value for all the results presented in the current work.  The resulting final least square estimator is thus:
\begin{equation}
M = F\cdot \widetilde{D}^\dag \cdot ( \widetilde{D}\cdot \widetilde{D}^\dag+\epsilon\mathds{I})^{-1} \cdot d
\label{eq:fit_LSQR_4}
\end{equation}which can be computed efficiently through standard numerical inversion techniques.
We emphasize that while computing the $[M]_{\beta k}$ optimal coefficients, there is no need for keeping the 3-center Coulomb integrals or the associated $\mathcal{F}^{V}_{\beta}(\alpha\alpha')$ coefficients: these can be computed once on-the-fly and discarded immediately, avoiding thus any extra memory consumption. In other terms, we never store explicitly the $F$ and $D$ matrices but only their $F\cdot D^\dag$ and $D\cdot D^\dag$ ($N_k\times N_k$) resulting products.

\subsection{Real-space grids generation }

In the present approach, the optimized $\lbrace {\bf r}_k \rbrace$ sets are generated for isolated atoms,  once for  every chemical species and their associated atomic basis sets. These atomic grids are then duplicated according to the molecule geometry to form the system-specific quadrature points. 
With Eq.~\ref{eq:fit_LSQR_4} defining the optimal $M$ for a given $\lbrace {\bf r}_k \rbrace$ set,   locations are adjusted  so as to minimize the fit error for a single atom test co-densities, using the Coulomb metric:
\begin{equation}
\argmin_{\{\mathbf{r}_k\}} \sum_{\rho}\Big|\Big| \mathcal{F}^{RS} (\rho)  - \mathcal{F}^{V} (\rho)  \Big|\Big|_V^2  \label{eq:fit_LSQR_5}
\end{equation} 
For the sake of keeping the optimization process relatively simple, we structured the $\lbrace {\bf r}_k \rbrace$ set over four different shells, each one replicated with a different number of radii. The only parameters of the optimization problem are thus the number of radii and their length. The four base shells taken here and denoted $A_1$, $A_2$, $A_3$ and $B_1$ are subsets of the Lebedev quadrature grids\cite{LEBEDEV1975}  (denoted here $L_i$ for the Lebedev grid of order i) in the sense that:
\[
\begin{split}
&L_{3\phantom{1}}=A_1 \\
&L_{5\phantom{1}}=A_1\cup A_2 \\
&L_{7\phantom{1}}=A_1\cup A_2\cup A_3 \\
&L_{11}=A_1\cup A_2\cup A_3\cup B_1 \\
\end{split}
\]The determination of the number of radii associated with each shell has been done through experimentation until satisfying configurations were obtained.  Base shells are provided in the SI Tables S4-7, and the resulting atomic quadrature grids for atoms H, C, N and O  in the SI Tables S8-11.\cite{suppinfo}  Optimizing freely all $\mathbf{r}_k$ locations and not only the radii may leads to significant improvement with respect to the grid size / accuracy ratio. Providing such grids falls however outside the scope of the present work.

\subsection{Technical details }

Benchmark Hartree-Fock and MP2 calculations were performed on a standard set of 28 medium size organic molecules containing unsaturated aliphatic and aromatic hydrocarbons or heterocycles, aldehydes, ketones, amides and nucleobases. Such a test set was originally proposed by  Thiel and coworkers \cite{Sch08} for reference optical excitations calculations within e.g. coupled cluster, \cite{Sch08,Sil10b} TD-DFT \cite{Jac09} and more recently Bethe-Salpeter \cite{Jac15,Bru15,Kra17} formalisms. We adopt   the MP2/6-31Gd geometries supplied in Ref.~\citenum{Sch08}. 

The assessment of the scaling properties of the present real-space quadrature RPA implementation is further performed on the oligoacenes family from benzene to hexacene using the B3LYP cc-pVTZ geometries available in Ref.~\citenum{Ran16}, complemented by the decacene, recently observed, \cite{Kru18} and the (hypothetical) octacene, both relaxed at the B3LYP/6-31Gd level. Finally, we consider the $C_{60}$ fullerene (B3LYP/6-311Gd geometry provided in the SI) and the octapeptide angiotensin II molecule originally proposed by Eshuis and coworkers. \cite{Esh10}

All calculations are performed with input molecular orbitals generated at the (spherical) cc-pVTZ \cite{CCPVTZ} Hartree-Fock level using the NWChem package.\cite{NWCHEM}  The corresponding (cartesian) cc-pVTZ-RI auxiliary basis \cite{Wei02} was adopted in all resolution-of-identity (RI) approaches (RI-SVS, RI-V and real-space quadrature RI-RS). For sake of comparison, Hartree-Fock exchange and MP2 correlation energies were calculated exactly, namely without any RI approximation, using the NWChem package as well.  All calculations are performed without any frozen-core approximation. 

The set $\{\rho\}$ of test co-densities can be adjusted depending on the needs, for example to match a specific subset of the wave functions co-densities. In the rest of this work, we adopt the following settings: 
\begin{equation}
\{\rho\}=(\{\alpha\}\otimes\{\alpha'\}_{l\leq 2})\cup\{\beta\}  
\label{eq:fit_LSQR_7}
\end{equation}
for both the single atom $\lbrace \mathbf{r}_k \rbrace$ sets problem and the full system optimization of $M$ coefficients. Limiting the second $\{\alpha' \}$ atomic orbital (AO) basis set to $s$, $p$ and $d$ orbitals allowed to speed up the computation while no significant change of accuracy was observed. In the minimization process, the weight on the $s$ and $p$ AO orbitals have also been stressed with respectively factors 4 and 2, so as to increase focus on  the low-order multipole charge and dipole component of the co-densities. Inclusion of the $\{\beta\}$ auxiliary orbitals within the test set slightly improves regularity of the errors. 

We adopt real-space $\lbrace {\bf r}_k \rbrace$ sets that contain typically 320 points per C, N and O atom, and 180 for hydrogen. This size corresponds  to about 3 times the size of the corresponding cc-pVTZ-RI auxiliary basis set. In the present study, we do not seek to look for minimal grid sizes, showing here below that excellent accuracy and small crossover between RI-RS and RI-V can already be obtained with such parameters. Details about the optimized real-space $\lbrace {\bf r}_k \rbrace$ sets, optimized for the cc-pVTZ and cc-pVTZ-RI Gaussian basis sets following Eq.~\ref{eq:fit_LSQR} and \ref{eq:fit_LSQR_5}, are provided  in the SI. \cite{suppinfo}

The Laplace transform (LT) RPA correlation energy calculations are based on time and frequency grids  described in the Appendix together with convergence tests for the benzene correlation energy, other molecules being reported in the SI.\cite{suppinfo} The present RI calculations, including the standard RI-V, RI-SVS and the newly developed real-space RI-RS, with and without Laplace Transform, are performed with a specific pilot code building on the Coulomb integral libraries implemented in the {\sc{Fiesta}} code. \cite{Jac15,Duchemin16,Jin16} 

\section{Results}

\subsection{Assessing the accuracy of the optimized real-space grid :  exact exchange  and  MP2 correlation energies}

As a first accuracy test of the present real-space RI implementation, namely to assess the quality of the co-density fits, we calculate both the  exact exchange  energy:
\begin{equation}
\begin{split}
E_{xx}=& -  \sum_{ij}^{occ} (ij|ij) 
\end{split}  \label{exxri}
\end{equation}written here for a spin compensated system, and the M{\o}ller-Plesset (MP2) correlation energy:
\begin{equation}
E_C^{MP2} = - \sum_{ij}^{occ}  \sum_{ab}^{virt}  { (ia|jb) [ 2(ia|jb) - (ib|ja)] \over \varepsilon_a + \varepsilon_b -  \varepsilon_i  -  \varepsilon_j }
\label{eq:E_MP2}
\end{equation}The exchange and MP2 energies are calculated using the RI expressions of the 2-electron integrals, namely e.g.:
\begin{equation}
(ia|jb) \overset{RI-X}{=} \sum_{\beta \beta'} \mathcal{F}^X_{\beta}(ia) V_{\beta \beta'}  \mathcal{F}^X_{{\beta}'}(jb)  \label{eq:RIX}\end{equation}where X=SVS, V or RS depending on the selected  scheme. Since we want to specifically address the accuracy of the fitting technique, we avoid at that stage using extra approximations such as Laplace transform techniques.

\begin{figure}
\includegraphics[width=10cm]{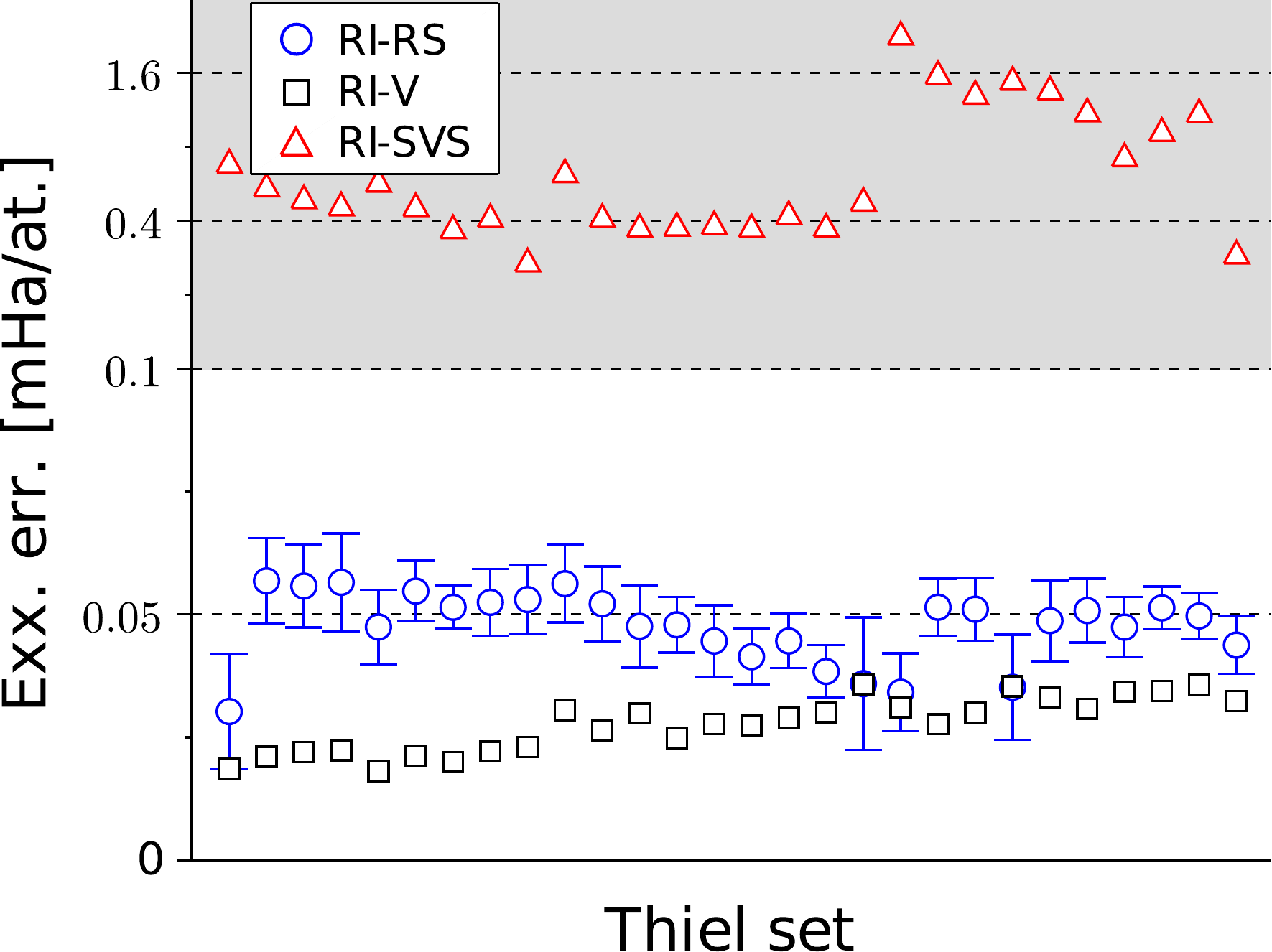}
\caption{Exchange energy (Exx) error as compared to exact calculations for various RI approximations (RI-RS, RI-V and RI-SVS). Errors are given in milli-Hartree ($mHa$) per atom (excluding H atoms). For RI-SVS data (red triangles) we adopt a log scale (grey shaded area). The error bar on the RI-RS data (blue circles) is related to the variance of the exchange energy error distribution ($mHa$ per atom, computed over 40 random orientations) with respect to molecule orientation (see text). The molecules are ordered from left to right following the original order provided in Ref.~\citenum{Sch08} and in the SI Table S1.}
\label{fig1}
\end{figure}

The results, namely the error as compared to exact calculations for the 28 Thiel's set molecules, are provided in Fig.~\ref{fig1} for exchange energy calculations. Clearly, the $\lbrace {\bf r}_k \rbrace$ sets adopted in this study provide errors that are of the same magnitude as the targeted RI-V approximation, and much smaller than those obtained with the RI-SVS scheme.   We recall that our real-space quadrature was optimized to reproduce the coefficients $\mathcal{F}_{\beta}^{V}$ of the RI-V formalism (see Eq.~\ref{eq:fit_LSQR}) so that it should not be  expected that the RI-RS approach yields errors smaller than the RI-V.   

The RI-RS results are provided with an "error bar" that represents the variance of the error distribution when the molecules are rotated. Contrary to the  $\lbrace \beta \rbrace$ Gaussian auxiliary basis, the atomic $\lbrace {\bf r}_k \rbrace$ set is not rotationally invariant. Clearly, however, such a variance remains marginal as compared e.g. to the differential of errors between the standard Coulomb-fitting (RI-V) and density fitting (RI-SVS) schemes.

\begin{figure}
\includegraphics[width=10cm]{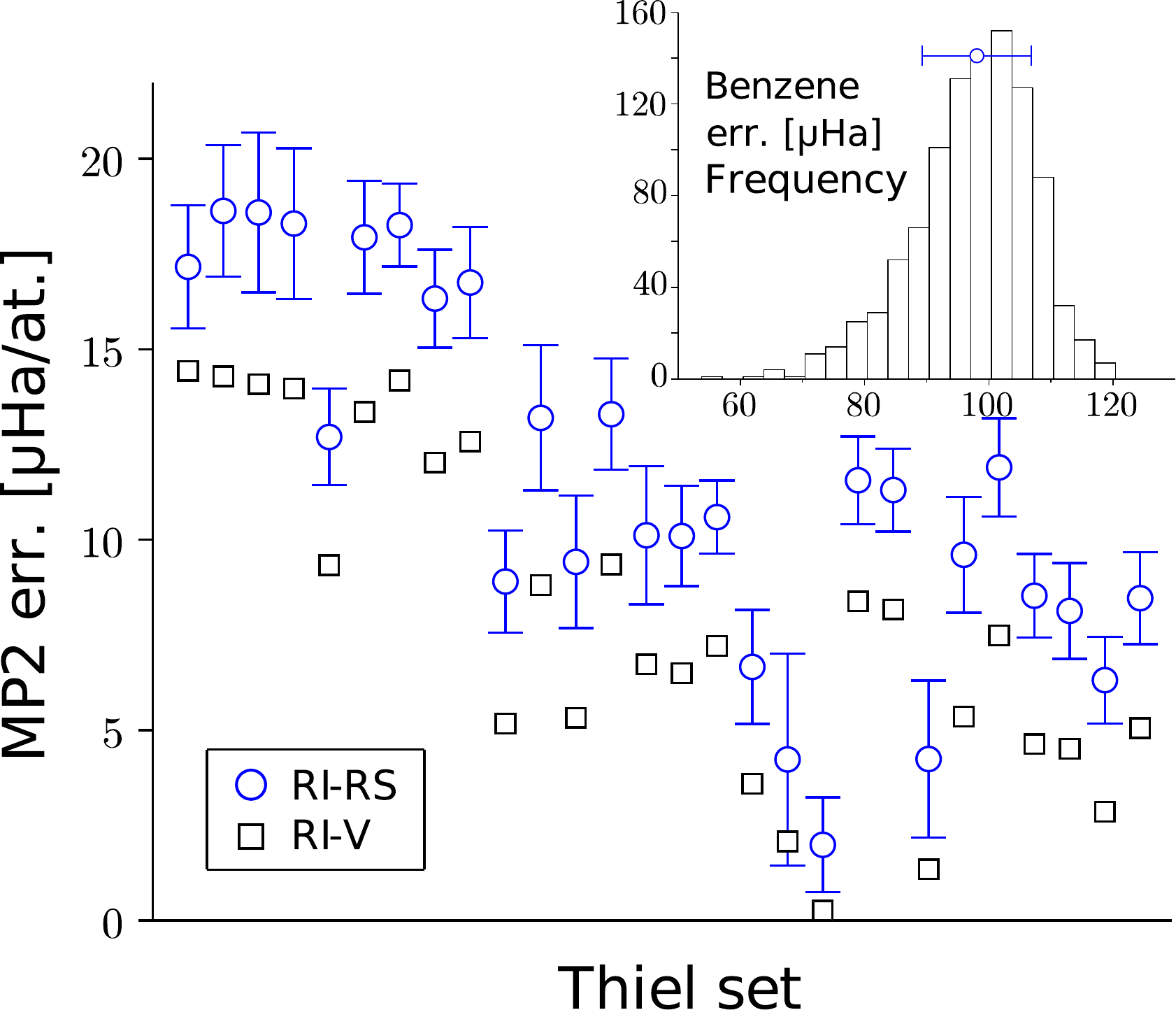}
\caption{MP2 correlation energy error as compared to exact calculations for the standard Coulomb-fitting (RI-V) and  real-space quadrature (RI-RS) approximations. Errors are given in micro-Hartree (${\mu}Ha$) per atom (excluding H atoms). The error bar on the RI-RS data (blue circles) is related to the variance of the energy error distribution (${\mu}Ha$ per atom) with respect to molecule orientation  (see text). Inset: details of benzene RI-RS MP2 correlation energy error (${\mu}Ha$) over a thousand random orientations with the corresponding variance (blue segment).  Reference values (no-RI) and errors are provided in SI Table S1. }     
\label{fig2}
\end{figure}

We now turn to the MP2 correlation energies (Fig.~\ref{fig2}). Again, we observe that the RI-RS quadrature does not degrade significantly the targeted RI-V MP2 correlation energies, with an error remaining lower than a few tens of $\mu$Hartree per atom. 
Similarly, the variance of the error remains small. We provide in the Inset of Fig.~\ref{fig2}  the actual distribution of errors obtained for a thousand independent random orientations of the benzene molecule, reporting the related variance error bar. We can conclude from the present set of benchmark calculations that the real-space representation (RI-RS), with the adopted $\lbrace {\bf r}_k \rbrace$ distributions size,
reproduce accurately the RI-V Coulomb integrals involving co-densities (products $\phi_i \phi_j$ in the exact exchange expression) and transition densities (products $\phi_i \phi_a$ in the MP2 energy formula) at the core of all explicitly correlated perturbative techniques. 

\subsection{Laplace transformed RPA}

We now turn to the central application of the present study, namely the calculation of the correlation energy within the random phase approximation (RPA). \cite{Pin52,Noz58,Lan77} We show in particular that the  real-space quadrature  RI-RS, combined with the standard Laplace transform (LT) approach, \cite{Alm91} allows reducing the scaling with system size down to $\mathcal{O}(N^3)$, instead of the  $\mathcal{O}(N^4)$ scaling of standard RI-RPA implementations,\cite{Esh10} without invoking any localization nor sparsity arguments.

Following seminal papers,\cite{Fur05,Fuc05,Fur08} we start with the adiabatic-connection fluctuation-dissipation theorem (ACFDT) formula to the RPA correlation energy : 
\begin{equation}
E^{RPA}_C = {1 \over 2\pi } \int_0^{\infty} d\omega \; Tr \large[ {\ln}(1-{{\chi}_0(i\omega)}\cdot v) + {{\chi}_0(i\omega)}\cdot v  \large]
\label{eq::ERPA}
\end{equation}where $v$ is the bare Coulomb operator and $\chi_0(i\omega)$ the independent-electron density-density susceptibility at imaginary frequency, that is for closed-shell systems:
\begin{eqnarray}
\chi_0({\bf r},{\bf r}' ; i\omega)  &= &  2 \sum_{ja} \frac{\phi _j^*({\bf r})  \phi _a({\bf r}) \phi _a^*({\bf r}')  \phi _j({\bf r}')  }{ i\omega - (\varepsilon_a - \varepsilon_j) } + cc   \label{eq:chi0_rr} \\
&  \overset{RI}{\simeq}  &\sum_{\beta\beta'} \beta(\mathbf{r})\beta'(\mathbf{r}') \left[ 2 \sum_{ja} \frac{\mathcal{F}_{\beta}(\phi _j\phi _a) \mathcal{F}_{\beta'}(\phi _a\phi _j) } {i\omega - (\varepsilon_a - \varepsilon_j) } + cc \right]  \label{eq:chi0_bb}\\
& \doteqdot & \sum_{\beta\beta'} \beta(\mathbf{r})\beta'(\mathbf{r}') \big[\chi_0^{RI}(i\omega)\big]_{\beta\beta'}
\end{eqnarray}with $(j,a)$ indexing (occupied/virtual) molecular eigenstates. The construction of the $\chi_0^{RI}(i\omega)$ matrix according to Eq.~\ref{eq:chi0_bb} clearly scales as $\mathcal{O}(N^4)$. 

\begin{figure}
\includegraphics[width=12cm]{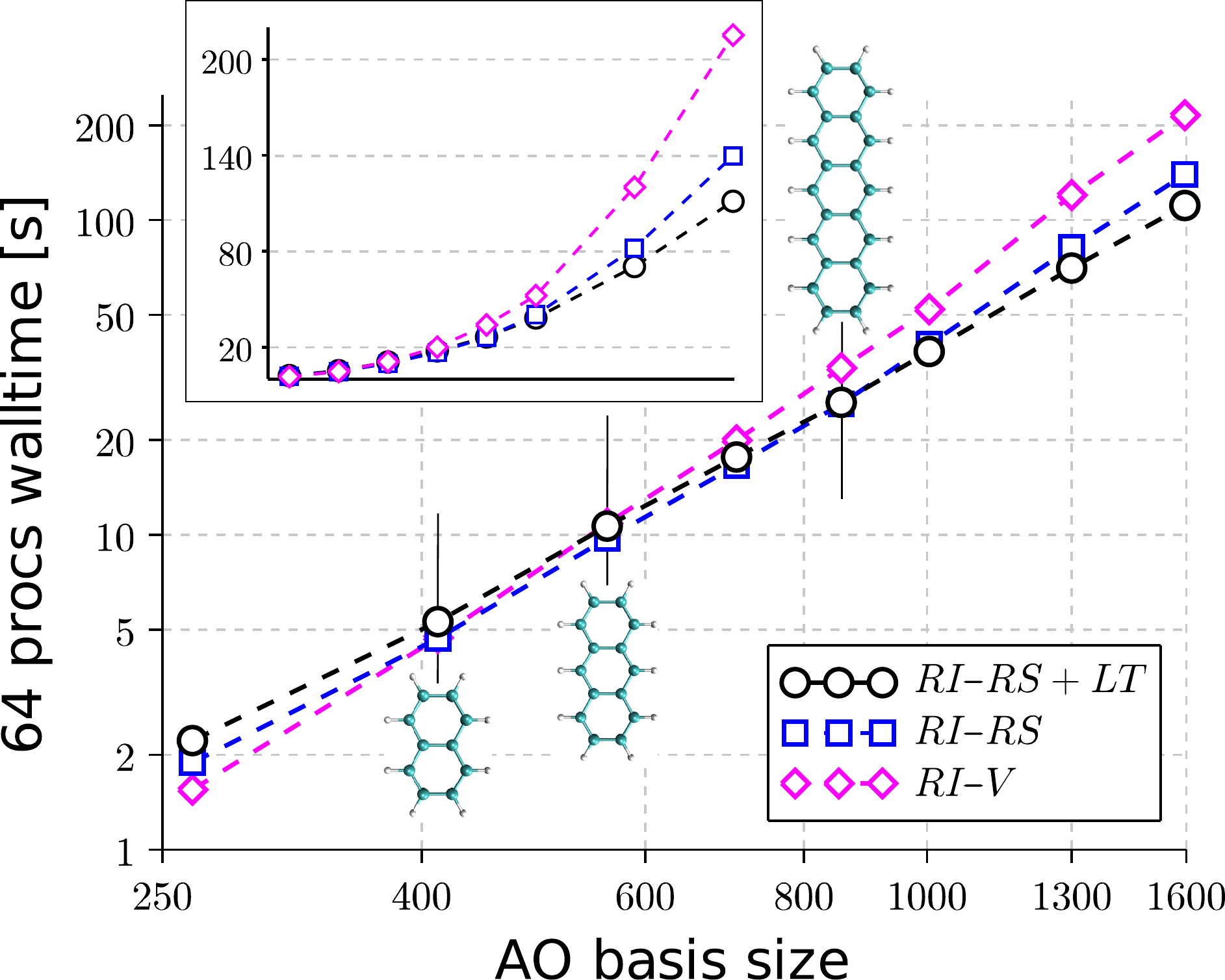}
\caption{Total walltime for the calculation of the RPA correlation energy over the acene family (benzene to decacene) using several RI schemes. We compare in particular the standard Coulomb fitting RI-V approach, the real-space quadrature RI-RS scheme, with and without Laplace transform (LT) technique. The abscissa provides the size of the (cc-pVTZ) AO basis used to expand the molecular orbitals. Both axis are displayed in log scale. The crossovers between the various RI formalisms are indicated by vertical short segments with the corresponding molecules. Inset : same data points without log scales. Walltimes are given for a run on 64 processors described in Note~\citenum{CPU-details}. }
\label{fig3}
\end{figure}

To discuss such scaling properties, we compare first in Fig.~\ref{fig3} the total computing time for calculating the RI-RPA correlation energy within the standard Coulomb-fitting approach (RI-V) and the novel RI-RS formalism, using the acene family   from benzene to decacene as a test set.  Calculations are performed using the cc-pVTZ AO and cc-pVTZ-RI auxiliary basis sets, together with a 12-point quadrature rule for the imaginary frequency axis integration (see Appendix). The corresponding correlation energies are provided in the Appendix for benzene, and in the SI for other acenes, demonstrating again the accuracy of the real-space approach as compared to the targeted RI-V approach. 

We observe that the RI-RS scheme walltime becomes smaller than that of our RI-V implementation for acenes larger than naphthalene. We emphasize however that at that stage, both RI-V and RI-RS techniques offer the very same $\mathcal{O}(N^4)$ scaling, differing only by the expression of the $\mathcal{F}_{\beta}(\phi _j\phi _a)$ coefficients. This crossover is related to the effort coming from the $\mathcal{O}(N^4)$ computation of the full $\mathcal{F}_{\beta}(\phi _j\phi _a)$ coefficients set. 
Without any assumption on the sparsity, the RI-RS scaling relies only on dense algebra techniques and can thus be implemented efficiently.

In order to reduce this scaling, one thus needs to avoid explicit calculation of the full $\mathcal{F}_{\beta}(\phi _j\phi _a)$ coefficients set. We achieve this by evaluating the independent-electron susceptibility directly in the real-space representation before transforming it back to the normal auxiliary basis representation:
\begin{equation}
\big[\chi_0^{RI}(i\omega)\big]_{\beta\beta'} = 
   \sum_{kk'}\;  M_{\beta k} \; \chi_0({\bf r}_k,{\bf r}_{k'} ; i\omega) \;  M_{\beta' k'}   \label{matmult}
\end{equation}The second step consists in applying the well known Laplace transform (LT) technique \cite{Alm91} so as to first compute $\chi_0({\bf r}_k,{\bf r}_{k'} ;i\tau)$ in the time domain where its expression is separable \cite{Roj95,Kal14} and transform it back to the frequency domain, using quadrature rules to form $\chi_0({\bf r}_k,{\bf r}_{k'} ;i\omega)$ (Eq.~\ref{chiOkkLT}). Such scheme allows us to  work with factorized expression of  $\chi_0({\bf r}_k,{\bf r}_{k'} ;i\tau)$ (Eq.~\ref{chiOkktau}):
\begin{eqnarray}
\label{chiOkkLT}
& & \chi_0({\bf r}_k,{\bf r}_{k'};i\omega) = \sum_{\tau} c_\tau(\omega) \chi_0({\bf r}_k,{\bf r}_{k'};i\tau)  \\
\label{chiOkktau}
& & \chi_0({\bf r}_k,{\bf r}_{k'};i\tau) = G^{<}({\bf r}_k,{\bf r}_{k'};i\tau) G^{>}({\bf r}_k,{\bf r}_{k'};-i\tau) 
\end{eqnarray}introducing the propagators of the occupied states  and of the unoccupied states, respectively:
\begin{eqnarray}
G^{<}({\bf r}_k,{\bf r}_{k'};i\tau) & = i &\sum_j \phi_j({\bf r}_k) \phi_j({\bf r}_{k'}) e^{\varepsilon_j \tau} \\
G^{>}({\bf r}_k,{\bf r}_{k'};-i\tau) & = -i &\sum_a \phi_a({\bf r}_k) \phi_a({\bf r}_{k'}) e^{-\varepsilon_a \tau} 
\end{eqnarray}with $\tau >0$ and the zero of occupied/virtual electronic energy levels taken at the Fermi level.  As a result of the decoupling of occupied and virtual states, the $G^{<}$ and $G^{>}$ propagators can be obtained with $\mathcal{O}(N^3)$ operations. Further, the entire $\big[\chi_0^{RI}(i\omega)\big]_{\beta\beta'} $ matrix stems from a combination of  Hadamard product (Eq.~\ref{chiOkktau}) and standard matrix operations (Eq.~\ref{matmult}-\ref{chiOkkLT}), yielding an overall $\mathcal{O}(N^3)$ process.

We can now report in Fig.~\ref{fig3} the full calculation walltime associated with the RI-RS+LT approach. Laplace transform quadratures for each of the 12 imaginary frequencies are performed with a  grid of 18 imaginary times, yielding  fully converged RPA energies (see Appendix). With such running parameters, the RI-RS+LT approach becomes more efficient than the standard RI-V formalism for systems larger than anthracene, outperforming the RI-RS approach for molecules larger than pentacene.

\begin{figure}
\includegraphics[width=12cm]{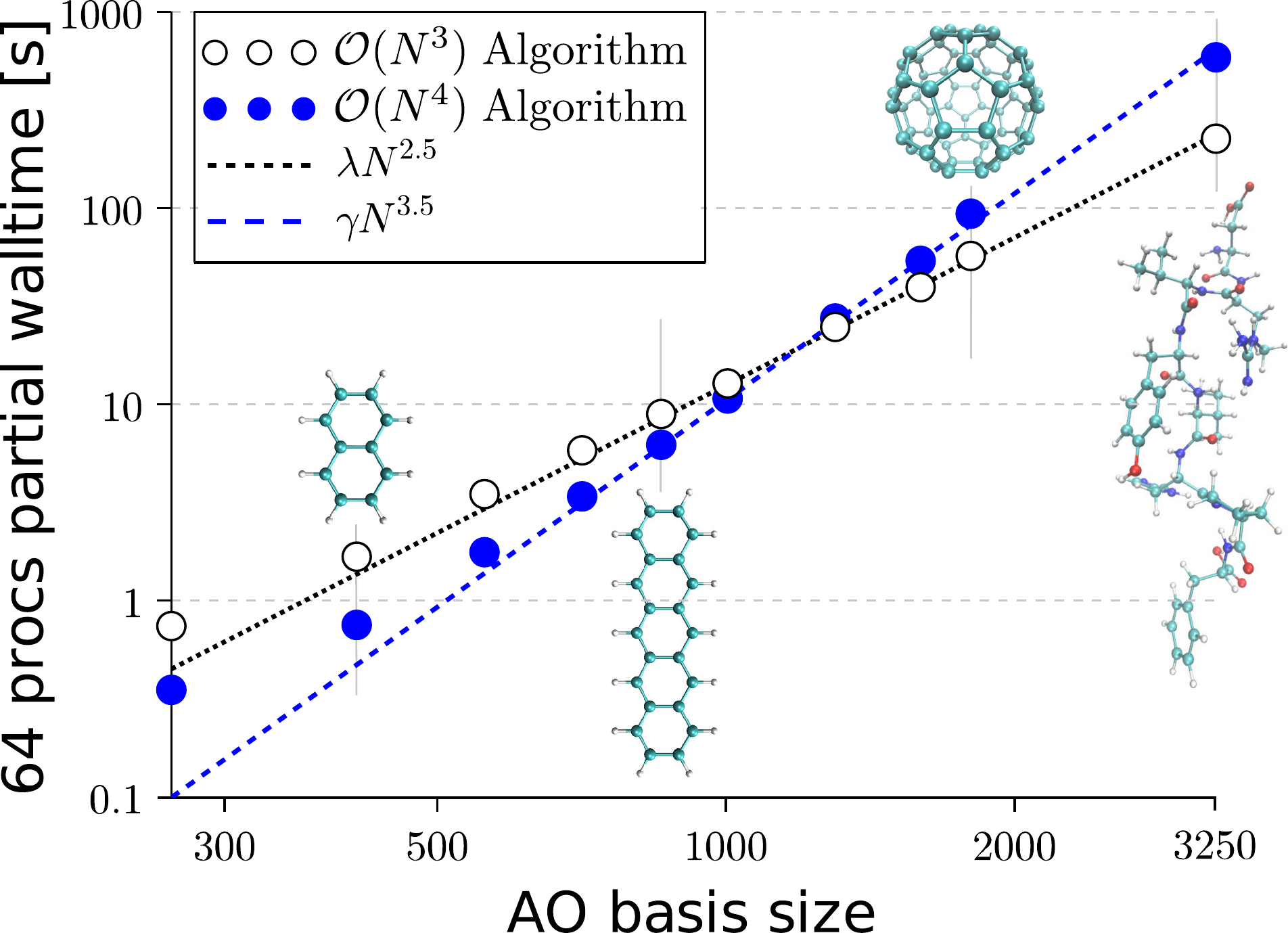}
\caption{Partial walltime associated with the RI-RS RPA correlation energy calculations with and without Laplace Transforms (LT), removing the extra cost associated with obtaining the $\mathcal{F}_{\beta}(\phi _j\phi _a)$ coefficients in the case of the no-LT approach.\cite{CPU-details} The test covers the acene family plus the $C_{60}$ fullerene and octapeptide angiotensin II molecule. The abscissa provides the size of the (cc-pVTZ) AO basis used to expand the molecular orbitals. Both axis are displayed in log scale. Dotted and dashed lines are a schematic guide to the eyes for scaling properties. Walltimes are given for a run on 64 processors described in Note~\citenum{CPU-details}. }
\label{fig4}
\end{figure}

To better assess the scaling properties associated with the resolution of Eq.~\ref{eq::ERPA} within standard and Laplace Transformed RI-RPA approaches, we single out the corresponding computation walltimes in Fig.~\ref{fig4}. We assume in particular that fitted co-densities coefficients are already available in the case of standard approaches, so that standard RI-V and RI-RS computational loads are equivalent. To probe larger systems,  the  test set is  extended with the $C_{60}$ fullerene and the original octapeptide angiotensin II molecule  proposed by Eshuis and coworkers.\cite{Esh10}

Without accounting for the fitting of co-densities calculation, the Laplace Transform RI-RS  RPA algorithm supersedes the standard RI-RS (or RI-V) RPA approach for systems larger than hexacene. This delayed crossover (as compared to Fig.~\ref{fig3}) can be imputed to both the Laplace transform overheads and the fact that with the ${\lbrace {\bf r}_k \rbrace}$ real space point set presently used, $\big[\chi_0(i\omega)\big]_{kk'}$ matrices are roughly $3\times 3$ times bigger than $\big[\chi_0(i\omega)\big]_{\beta\beta'}$ ones, leading to a $3^3$ prefactor in the linear algebra operations. This last fact demonstrates the importance of operating on small  ${\lbrace {\bf r}_k \rbrace}$ quadrature sets. The additional cost coming from obtaining the fitting parameters $\mathcal{F}_{\beta}(\phi _j\phi _a)$  only adds to the cost of the no-Laplace-Transform RI approaches,  bringing the overall crossover at the level of a pentacene molecule, as exemplified in Fig.~\ref{fig4}. We finally observe $\mathcal{O}(N^{3.5})$ and $\mathcal{O}(N^{2.5})$ scaling laws (see dotted lines), indicating that the expected asymptotic $\mathcal{O}(N^{4})$ and $\mathcal{O}(N^{3})$ behaviours are not yet reached for the tested molecule sizes. We conclude this section by mentioning that the total walltime for the (RI-RS+LT) calculation of the cc-pVTZ RPA correlation energy for $C_{60}$ takes less than 6500 secs on a single processor.  

\section{Discussion}

The present implementation can be compared to the real-space-grid imaginary-time approach introduced originally in the framework of $GW$ calculations by Rojas and coworkers \cite{Roj95} or the recent real-space-grid imaginary-time RPA implementation by Kaltak and coworkers. \cite{Kal14} In such studies, the real-space grid was obtained as the Fourier transform of the planewave basis used to expend the Bloch states in a pseudopotential or PAW framework for periodic systems. Alternatively, the work of Moussa \cite{Mou14} demonstrated as well cubic scaling for a random-phase approximation with second-order screened exchange formalism, exploiting a real-space grid for both the primary and auxiliary basis sets, combined with nested low-rank approximations to energy denominators.  The present RI-RS approach, while targeting all-electron atomic-basis calculations,  preserves the use of the standard representation of molecular orbitals, related co-densities and response operators in terms of atomic orbitals and their associated  auxiliary bases, adopting further standard Laplace transform techniques. 

A central issue in real-space representations, in particular when performing all-electron calculations, concerns the size of the real-space grid that strongly influences the crossover with standard RI implementations and the memory requirements. This is all the more important in the present study since we aim to calculate and store intermediate non-local operators such as the ${\chi}^0({\bf r}_k, {\bf r}_{k'}; i\tau)$ susceptibilities, and not only local functions such as the charge density or the DFT exchange-correlation potential and energy density. Since our real-space $\lbrace {\bf r}_k \rbrace$ distribution must serve in a quadrature reproducing co-densities involving   molecular orbital products, one may expect that it should be as large as standard grids \cite{Gil93} used to represent the charge density in DFT codes. Taking as  an example the Gaussian09 code, the default  DFT grid involves  about 7000 grid points per atom after pruning. This is consistent with the recommended "Grid3" in the original paper by Trutler and Alrichs \cite{Tre95}  yielding 5980 pruned points for elements from Li to Ne and that serves as the default in Turbomole. 

Such standard DFT grid sizes are much larger than the number of real-space points used in the present RS approach, roughly 180 for Hydrogen and 320 per non-H atom (C, N, O). The present $\lbrace {\bf r}_k \rbrace$ sets were optimized so that the RI-RS scheme faithfully reproduces the standard RI-V density fitting results.  Each set is composed of a number of different shells of high symmetry points, each one associated with a set of different radii. This minimization process results in a non-uniform ${\lbrace {\bf r}_k \rbrace }$ distribution of real-space points. As emphasized above, we did not seek to explore here in great details the set size to accuracy ratio, for our goal was to demonstrate that one can find an accurate real-space representation yielding a crossover with standard techniques for small system sizes. Smaller optimized sets of real-space points may certainly be explored in the future. 

Another advantage the RI-RS+LT RPA scheme lies in the related $\mathcal{O}(N^2)$ memory footprint associated with the underlying matrix algebra. In the case of the $C_{60}$ molecule, the relevant sizes are 1800 spherical AOs, 6060 cartesian auxiliary orbitals, and 19140 real-space quadrature points. On a single processor run, the set up of the RI-RS (Eq.~\ref{eq:fit_LSQR}) peaks at about 8~Gb of memory, while leaving at exit a memory footprint below 1Gb for the $M_{{\beta}k}$ coefficients. Further, each $\chi_0({\bf r}_k,{\bf r}_{k'};i\tau)$ requires about 3 Gb. In regards of a parallelization scheme oriented towards CPU efficiency, one can benefit from storing all the ($n\tau$) $\chi_0({\bf r}_k,{\bf r}_{k'};i\tau)$ in memory. 

As emphasized here above, the present cubic scaling in terms of floating point operations, and quadratic scaling in terms of memory load, was obtained without invoking localization nor sparsity considerations. In particular, our approach does not require the use of the density-fitting (RI-SVS) approach with its sparse 3-center overlap matrix tensor. However, another class of localization properties, based on the exponential decay in real-space of the one-body Green's function in gaped systems,\cite{Sch00} can be easily combined with the present  approach. These localization properties, that strongly depend on the electronic properties of the system of interest, are reminiscent of the low-scaling techniques based on local AO formulations in the treatment of MP2  \cite{Has93,Aya99}  or RPA \cite{Kal15,Sch16,Lue17} correlation energies. \cite{notegreens}   Such additional considerations, together with the  stochastic approach by Neuhauser and coworkers, \cite{Neu13}   may be easily combined and explored in the future to further reduce the memory and computing time. 

\section{ Conclusion }

We have introduced a separable resolution-of-the-identity based on a real-space quadrature of co-densities. The efficiency of our approach relies on setting up an optimal and compact distribution of real-space points $\lbrace {\bf r}_k \rbrace$ allowing excellent accuracy, as exemplified in the case of the exact exchange energy and further the MP2 and RPA correlation energies, taking as a test case a large set of molecular systems. Our approach preserves the use of standard Gaussian atomic orbitals and related auxiliary basis sets for all-electron calculations, the real-space set of points being used as an  intermediate representation. We demonstrate that such an approach leads to calculating RPA correlation energies with a cubic  scaling in terms of operations, quadratic in memory, without invoking any localization nor sparsity considerations that may be combined in the future. The limited number of needed real-space points allows early crossovers with traditional Coulomb-fitting RI-RPA calculations for systems as small as naphthalene or anthracene (Fig.~\ref{fig3}). The application of such a real-space separable RI to other explicitly correlated techniques, such as the $GW$ and Bethe-Salpeter equation (BSE) formalisms for calculating charged and neutral electronic excitations in molecular systems,\cite{Bla18} are currently under exploration.

\begin{acknowledgements}

The authors thank Thierry Deutsch and Pierre-Francois Loos for their critical reading of the manuscript and Denis Jacquemin for running the benzene RPA calculations with the Turbomole code. This research used resources from the French GENCI supercomputing centers under project no. A0030910016. 

\end{acknowledgements}

\appendix*

\section{Imaginary frequency and time quadratures}

\begin{figure}
\includegraphics[width=8cm]{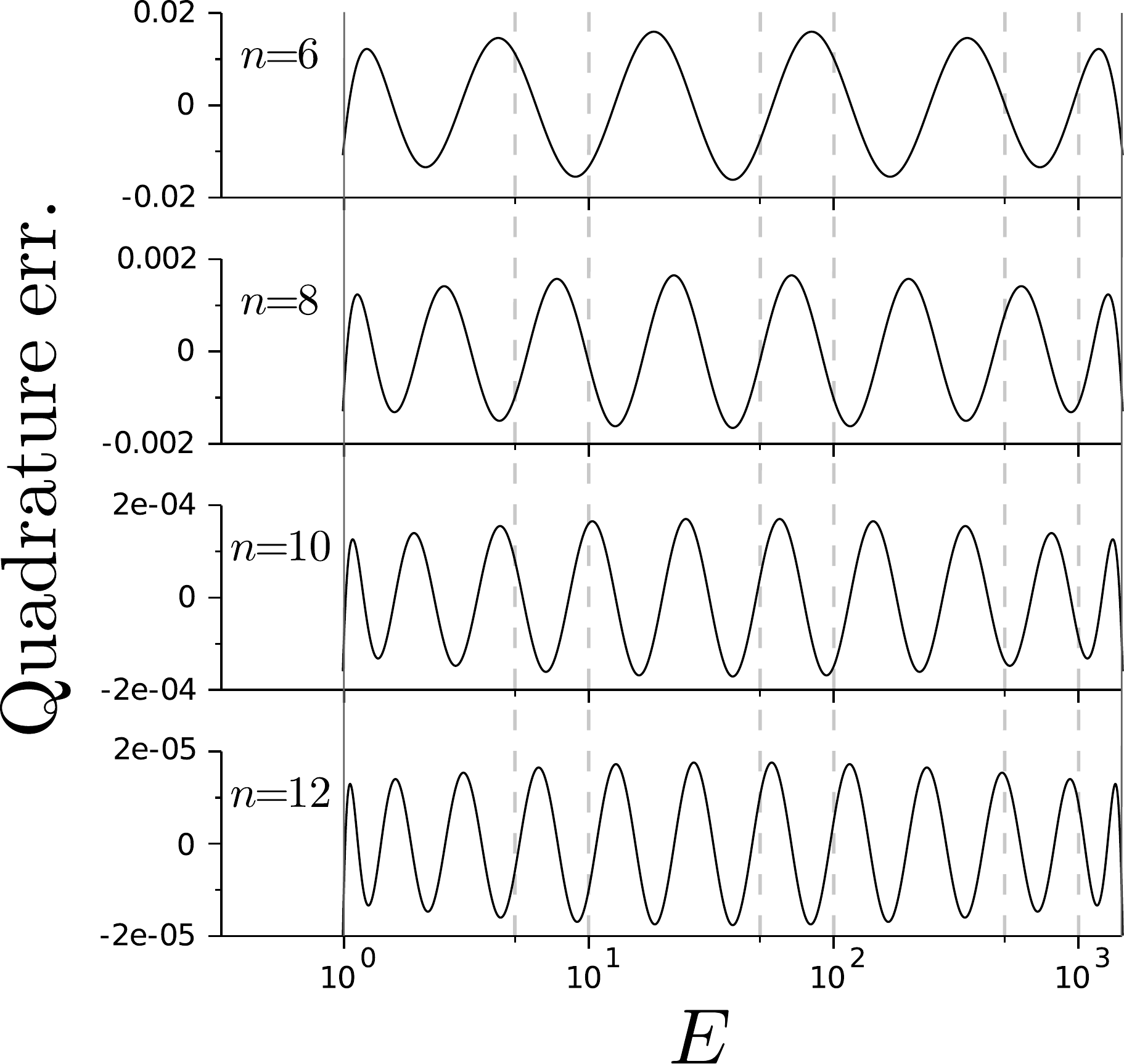}
\caption{Imaginary axis quadrature error (see Eq.~\ref{eq::quad1}) for  $n\omega\in\{6,8,10,12\}$, $E_{min}=1$ and $E_{max}=1500$}
\label{fig5}
\end{figure}

We briefly outline the imaginary frequency and time  quadrature strategies adopted for calculating the RPA correlation energy and for setting the Laplace transform.  Relying on the poles structure of the susceptibility (Eq.~\ref{eq:chi0_rr}), we seek for a quadrature that reproduces the contribution of all possible poles in the imaginary axis integral, using the exact relation:
\begin{equation}
\int_{0}^{\infty} d\omega \Big[  \frac{1}{E-i\omega} + \frac{1}{E+i\omega} \Big] = \pi 
\label{eq::quad1}
\end{equation}
where the pole energy $E$ can vary from $E_{min}$, i.e. the electronic energy  gap, to $E_{max}$ determined by the maximum transition energy between occupied and virtual energy levels. Even if an optimal solution of this problem could be in principle determined through the minimax fitting approach usually applied to Laplace transforms,\cite{Kal14} we prefer here a somewhat numerically simpler least square formulation, namely we seek for $n{\omega}$ frequencies $z_k$ and weights $w_k$ corresponding to:
$$
\argmin_{w_k, z_k} \left[ \int_{\ln (E_{min})}^{\ln (E_{max})} du \; \bigg|\bigg| \sum_k w_k \Big[  \frac{1}{e^u - iz_k} + \frac{1}{e^u+iz_k}   \Big] - \pi \bigg|\bigg|^2 \right]
$$Here, the log scale has been preferred so as to obtain uniform oscillations of the error as a function of the integration variable (see Fig.~\ref{fig5}), since integrating directly over the energy would favor larger errors in the small energy range.\cite{Kal14} Under such conditions, minimax and least square approaches should provide consistent results. 
The $z_k$ frequencies and their corresponding $w_k$ weights are used for the integration reported in Eq.~\ref{eq::ERPA}. Convergence tests are provided below in Table~\ref{tbl:example}  for benzene, other acenes being dealt with in the Supporting Information (SI).\cite{suppinfo}   Concerning benzene, our reference RI-V RPA correlation energy falls within less than 8 meV/molecule (0.19 kcal/mol) as compared to that obtained with other codes \cite{molgw,TURBOMOLE} with identified differences in the treatment of preceding Hartree-Fock calculations (RI \textit{vs} no-RI)  and use of the auxiliary basis (cartesian \textit{vs} spherical). Together with potential differences in the treatment of the energy integration, these variations result in very small energy differences (see SI Table S2 for details).\cite{suppinfo}    

For the Laplace transform, the  time grid $\lbrace \tau_p , p\in[1,n\tau] \rbrace$ is set accordingly so as to minimize the Laplace transform errors for the specific quadrature $(z_k,w_k)$ points obtained above, namely: 

$$
\argmin_{w_k^{p}, \tau_p} \left[ \sum_k \int_{\ln (E_{min})}^{\ln (E_{max})} du \; \bigg|\bigg| \sum_p w_k^{p} e^{-\tau_p\, e^u} -\Big[  \frac{1}{e^u - iz_k} + \frac{1}{e^u+iz_k}   \Big] \bigg|\bigg|^2 \right]
$$
where the weights $\lbrace w_k^{p} \rbrace$ depend on the targeted $z_k$ frequency. Again, the log scale is used so as to allow a regular sampling of the error oscillations. We draw the reader attention on the fact that the least square approach is clearly at advantage over the minimax fitting approach here since defining an alternant is not possible in the case of simultaneous fits (one fit for each $z_k$ frequency). The results provided here below for the benzene RPA correlation energy demonstrates already $<10^{-4}$Ha convergence for grids containing as few as $n\omega=6$ imaginary frequencies and $n\tau=9$ times. For $n\omega=12$ and $n\tau=18$, the parameters used in this manuscript for actual calculations, the correlation energy is clearly converged. 

\begin{table}
  \caption{RPA correlation energy (Hartree) for the benzene molecule. The various RI schemes are compared. Calculations are performed at the (spherical) cc-pVTZ level with the corresponding (cartesian) cc-pVTZ-RI auxiliary basis. The number of imaginary frequency in the RPA integration is given by n$\omega$. The number of times in the Laplace transform (LT) approach is given by n$\tau$.}
  \label{tbl:example}
  \begin{tabular}{cccccc}
    \hline
\multirow{2}{*}{n$\omega$}    & \multirow{ 2}{*}{RI-SVS}   &  \multirow{ 2}{*}{RI-V} &  \multirow{ 2}{*}{RI-RS} &  RI-RS+LT &  RI-RS+LT   \\
      &   &   &   &   (n$\tau$=1.5$\times$n$\omega$) &  (n$\tau$=2$\times$n$\omega$)  \\
    \hline
6 &     -1.25169968  & -1.25148948   & -1.25143651   & -1.25148267   & -1.25143097   \\
  8&  -1.25171657   & -1.25150621   & -1.25145324 & -1.25145455 & -1.25145327  \\
  10 &       -1.25171791  & -1.25150754  &  -1.25145456 &  -1.25145456 & -1.25145456 \\
    12 &     -1.25171791  & -1.25150754   & -1.25145456   & -1.25145456  &   -1.25145456  \\
    \hline
  \end{tabular}
\end{table}

\bibliography{manuscript}

\begin{thebibliography}{61}
\bibitem [{CPU()}]{CPU-details}%
  \BibitemOpen
  \href@noop {} {}\bibinfo {note} {All calculations were performed on 2 nodes
  of 2x16 cores of Intel® Xeon® Gold 6130 CPU @ 2.10GHz, with Intel®
  Omni-Path Interconnect, except for the RI-V RPA correlation energy of the
  octapeptide angiotensin II molecule where 4 half filled nodes were used in
  order to circumvent memory issues.}\BibitemShut {Stop}%
\end{thebibliography}

\end{document}


\setlength{\tabcolsep}{8pt}

\title{ Supporting information: Separable Resolution-of-the-Identity with All-Electron Gaussian Bases: Application to Cubic-scaling RPA}

\author{Ivan Duchemin}
\affiliation{Univ. Grenoble Alpes, CEA, INAC-MEM-L\_Sim, 38000 Grenoble, France}
\email{ivan.duchemin@cea.fr}
\author{Xavier Blase}
\affiliation{Univ. Grenoble Alpes, CNRS, Inst NEEL, F-38042 Grenoble, France}



\maketitle

In Section \ref{secSI1}, we provide (a) the MP2 correlation energies (Table~\ref{tbl:mp2SI}), (b) the RPA correlation energy for benzene as obtained with the present code, Turbomole and MolGW  (Table~\ref{tbl:benzeneSI}), (c) the RPA correlation energy for acenes, $C_{60}$ and the octapeptide angiotensin II molecule (Table~\ref{tbl:rpaSI}). In Section \ref{secSI2}, the set of $\lbrace {\bf r}_k \rbrace$ real-space points for H,C,N and O species is described. Finally we provide in Section \ref{secSI3}
the used geometries for the octacene and decacene (B3LYP 6-31Gd) and the $C_{60}$ fullerene (B3LYP 6-311Gd).
We finally demonstrate the $\mathcal{O}(N^3)$ construction of the LS-THC estimator\cite{Parrish12} starting from the present real-space RI formulation.

\section{MP2 and RPA correlation energies}
\label{secSI1}

\begin{center}
\LTcapwidth=\textwidth
\begin{longtable}{lllr}
  \caption{MP2 correlation energy (Hartree) for Thiel's set of molecules.\cite{Sch08} The various RI schemes are compared. Calculations are performed at the (spherical) cc-pVTZ level with the corresponding (cartesian) cc-pVTZ-RI auxiliary basis. The total reference energy (no RI) is provided while the RI results are given as an error as compared to the reference. For the real-space (RI-RS) scheme, we provide the mean signed error (MSE) averaged over 40 random orientations of the corresponding molecules (see main manuscript).  }
\label{tbl:mp2SI}
\\
\hline
\endfirsthead
\multicolumn{4}{c}%
{\tablename\ \thetable\ -- \textit{Continued from previous page}} \\
\hline
    &  ref. (no RI)\cite{NWCHEM} & RI-V err. & RI-RS MSE \\
\hline
\endhead
\hline \multicolumn{4}{r}{\textit{Continued on next page}} \\
\endfoot
\hline
\endlastfoot
        &  ref. (no RI)\cite{NWCHEM} & RI-V err. & RI-RS MSE \\
\hline
Ethene          &   -0.36621712 &     2.8864e-05 &     2.6618e-05 \\ 
Butadiene       &   -0.71282637 &     5.7159e-05 &     5.3500e-05 \\ 
Hexatriene      &   -1.06131929 &     8.4513e-05 &     8.8470e-05 \\ 
Octatetraene    &   -1.41053465 &     1.1173e-04 &     1.0533e-04 \\ 
Cyclopropene    &   -0.53351982 &     2.7989e-05 &     2.9524e-05 \\ 
Cyclopentadiene &   -0.88355154 &     6.6776e-05 &     7.6226e-05 \\ 
Norbornadiene   &   -1.25189681 &     9.9161e-05 &     1.1061e-04 \\ 
Benzene         &   -1.04326862 &     7.2164e-05 &     7.7232e-05 \\ 
Naphthalene     &   -1.72711598 &     1.2577e-04 &     1.3562e-04 \\ 
Furan           &   -0.94557368 &     2.5866e-05 &     3.3693e-05 \\ 
Pyrrole         &   -0.92636075 &     4.4035e-05 &     4.7447e-05 \\ 
Imidazole       &   -0.95859577 &     2.6609e-05 &     2.8459e-05 \\ 
Pyridine        &   -1.07638112 &     5.6094e-05 &     5.9644e-05 \\ 
Pyrazine        &   -1.11113492 &     4.0363e-05 &     3.4864e-05 \\ 
Pyrimidine      &   -1.10577313 &     3.8956e-05 &     4.4471e-05 \\ 
Pyridazine      &   -1.11510376 &     4.3281e-05 &     5.0146e-05 \\ 
Triazine        &   -1.13228003 &     2.1557e-05 &     1.5914e-05 \\ 
Tetrazine       &   -1.18621363 &     1.2456e-05 &    -3.3183e-06 \\ 
Formaldehyde    &   -0.42515595 &     5.2684e-07 &    -3.4332e-06 \\ 
Acetone         &   -0.80235559 &     3.3520e-05 &     3.5088e-05 \\ 
Benzoquinone    &   -1.49609284 &     6.5378e-05 &     7.1006e-05 \\ 
Formamide       &   -0.64873853 &     4.0474e-06 &    -7.3104e-06 \\ 
Acetamide       &   -0.83678121 &     2.1426e-05 &     2.8016e-05 \\ 
Propanamide     &   -1.02751877 &     3.7440e-05 &     4.5876e-05 \\ 
Cytosine        &   -1.57264992 &     3.7081e-05 &     4.4385e-05 \\ 
Thymine         &   -1.78790406 &     4.0576e-05 &     4.9361e-05 \\ 
Uracil          &   -1.59383066 &     2.2851e-05 &     2.8059e-05 \\ 
Adenine         &   -1.92267910 &     5.0528e-05 &     6.1021e-05 \\
\end{longtable}
\end{center}

\begin{center}
\LTcapwidth=\textwidth
\begin{longtable}{clll}
\caption{RPA correlation energy (Hartree) for the benzene molecule as obtained using the present implementation, together with the MOLGW and Turbomole codes. Available details about the calculations are provided. All calculations are performed at the (spherical) cc-pVTZ level with (spherical or cartesian) cc-pVTZ-RI auxiliary basis. }
\label{tbl:benzeneSI}
\\
\hline
    \multicolumn{1}{c}{energy}   
  & \multicolumn{1}{c}{code}
  & \multicolumn{1}{c}{Hartree-Fock}
  & \multicolumn{1}{c}{cc-pVTZ-RI  basis}  \\
\hline
\endfirsthead
\multicolumn{4}{c}%
{\tablename\ \thetable\ -- \textit{Continued from previous page}} \\
\hline
    \multicolumn{1}{c}{energy}   
  & \multicolumn{1}{c}{code}
  & \multicolumn{1}{c}{Hartree-Fock}
  & \multicolumn{1}{c}{cc-pVTZ-RI  basis}  \\
\hline
\endhead
\hline \multicolumn{4}{r}{\textit{Continued on next page}} \\
\endfoot
\hline
\endlastfoot
  -1.25150754 & present                   & no-RI &  cartesian \\
  -1.25147833 & MOLGW\cite{MOLGW}         & RI-V  &  spherical \\
  -1.25122935 & Turbomole\cite{TURBOMOLE} & RI-V  &  spherical \\
\end{longtable}
\end{center}

\begin{center}
\LTcapwidth=\textwidth
\begin{longtable}{lrrrrr}
\caption{RPA correlation energy (Hartree) for acenes, $C_{60}$ and the octapeptide angiotensin II molecule. The various RI schemes are compared. Calculations are performed at the (spherical) cc-pVTZ level with the corresponding (cartesian) cc-pVTZ-RI auxiliary basis. The number of imaginary frequencies in the RPA integration is given by n$\omega$=12. The number of times in the Laplace transform (LT) approach is given by n$\tau$.}
  \label{tbl:rpaSI}
\\
\hline
      & \multicolumn{1}{c}{\multirow{2}{*}{RI-SVS}}  
      & \multicolumn{1}{c}{\multirow{2}{*}{RI-V}}
      & \multicolumn{1}{c}{\multirow{2}{*}{RI-RS}}
      & \multicolumn{1}{c}{RI-RS+LT} 
      & \multicolumn{1}{c}{RI-RS+LT}   \\
      &   &   &   
      & \multicolumn{1}{c}{(n$\tau$=1.5$\times$n$\omega$)}
      & \multicolumn{1}{c}{(n$\tau$=2$\times$n$\omega$)}  \\
\hline
\endfirsthead
\multicolumn{6}{c}%
{\tablename\ \thetable\ -- \textit{Continued from previous page}} \\
\hline
      & \multirow{2}{*}{RI-SVS}   &  \multirow{2}{*}{RI-V} &  \multirow{ 2}{*}{RI-RS} &  RI-RS+LT &  RI-RS+LT   \\
      &   &   &   &   (n$\tau$=1.5$\times$n$\omega$) &  (n$\tau$=2$\times$n$\omega$)  \\
\hline
\endhead
\hline \multicolumn{6}{r}{\textit{Continued on next page}} \\
\endfoot
\hline
\endlastfoot
     Benzene  &  -1.25171791  & -1.25150754  & -1.25145456   & -1.25145456  &   -1.25145456  \\
     Naphthalene  & -2.04585075 & -2.04552127  & -2.04541965   & -2.04541965  &  -2.04541965  \\
    Anthracene     & -2.84161803 & -2.84117439 & -2.84103203 & -2.84103203 & -2.84103203 \\
    Tetracene         & -3.63813465 & -3.63757585 & -3.63740533 & -3.63740528 & -3.63740533 \\
    Pentacene        & -4.43569083 & -4.43501826 &  -4.43481146 & -4.43481136 & -4.43481146 \\
    Hexacene       & -5.23386340 &  -5.23307561 & -5.23284107 & -5.23284090 & -5.23284108  \\
    C60                    &   & -11.25611147 & -11.25546768 & -11.25546749 & -11.25546767 \\
    Octapeptide     &  & -17.53746478 & -17.53632073 & -17.53632074 & -17.53632073 \\
\end{longtable}
\end{center}

\section{Points sets for H, C, N and O species}
\label{secSI2}

We present the set of real-space $\lbrace {\bf r}_k \rbrace$ points generated to reproduce within the present RI-RS scheme the Coulomb-fitting RI-V data at the cc-pVTZ/cc-pVTZ-RI level.
Points set are generated as a combination of four different base shells associated with different radii. 
We start by providing the four base shells that we denote $A_1$, $A_2$, $A_3$ and $B_1$ (S4-7). These shells were constructed as subsets of the Lebedev quadrature grids\cite{LEBEDEV1975}  (denoted here $L_i$ for the Lebedev grid of order i) in the sense that:
\[
\begin{split}
&L_{3\phantom{1}}=A_1 \\
&L_{5\phantom{1}}=A_1\cup A_2 \\
&L_{7\phantom{1}}=A_1\cup A_2\cup A_3 \\
&L_{11}=A_1\cup A_2\cup A_3\cup B_1 \\
\end{split}
\]The base shells points are located on the unit sphere, while the associated radii are provided in atomic units. As indicated below, the origin \((0.0, 0.0, 0.0)\) point was also added on top of the different shell/radii combination for each species quadrature grid. The atomic quadrature grids for atoms H, C, N and O are reported in tables S8-11.

\begin{center}
\LTcapwidth=\textwidth
\begin{longtable}{rrr}
  \caption{Lebedev point subset 1/4.}
\label{tbl:LebSI1}
\\
\hline
       \multicolumn{3}{c}{set $A_1$} \\
  \hline
\endfirsthead
\multicolumn{3}{c}%
{\tablename\ \thetable\ -- \textit{Continued from previous page}} \\
  \hline
       \multicolumn{3}{c}{set $A_1$} \\
  \hline
\endhead
\hline \multicolumn{3}{r}{\textit{Continued on next page}} \\
\endfoot
\hline
\endlastfoot
       1.0000000000000000   &    0.0000000000000000   &    0.0000000000000001 \\
      -1.0000000000000000   &    0.0000000000000001   &    0.0000000000000001\\
       0.0000000000000001   &    1.0000000000000000   &    0.0000000000000001\\
       0.0000000000000001   &   -1.0000000000000000   &    0.0000000000000001\\
       0.0000000000000000   &    0.0000000000000000   &    1.0000000000000000\\
       0.0000000000000000   &    0.0000000000000001   &   -1.0000000000000000\\
\end{longtable}
\end{center}

\begin{center}
\LTcapwidth=\textwidth
\begin{longtable}{rrr}
  \caption{Lebedev point subset 2/4.}
\label{tbl:LebSI2}
\\
\hline
       \multicolumn{3}{c}{set $A_2$} \\
  \hline
\endfirsthead
\multicolumn{3}{c}%
{\tablename\ \thetable\ -- \textit{Continued from previous page}} \\
  \hline
       \multicolumn{3}{c}{set $A_2$} \\
  \hline
\endhead
\hline \multicolumn{3}{r}{\textit{Continued on next page}} \\
\endfoot
\hline
\endlastfoot
       0.5773502691896258   &    0.5773502691896257   &    0.5773502691896258\\
       0.5773502691896258   &    0.5773502691896257   &   -0.5773502691896257\\
       0.5773502691896258   &   -0.5773502691896257   &    0.5773502691896258\\
       0.5773502691896258   &   -0.5773502691896257   &   -0.5773502691896257\\
      -0.5773502691896257   &    0.5773502691896258   &    0.5773502691896258\\
      -0.5773502691896257   &    0.5773502691896258   &   -0.5773502691896257\\
      -0.5773502691896257   &   -0.5773502691896258   &    0.5773502691896258\\
      -0.5773502691896257   &   -0.5773502691896258   &   -0.5773502691896257\\
\end{longtable}
\end{center}

\newpage

\begin{center}
\LTcapwidth=\textwidth
\begin{longtable}{rrr}
  \caption{Lebedev point subset 3/4.}
\label{tbl:LebSI3}
\\
\hline
       \multicolumn{3}{c}{set $A_3$} \\
  \hline
\endfirsthead
\multicolumn{3}{c}%
{\tablename\ \thetable\ -- \textit{Continued from previous page}} \\
  \hline
       \multicolumn{3}{c}{set $A_3$} \\
  \hline
\endhead
\hline \multicolumn{3}{r}{\textit{Continued on next page}} \\
\endfoot
\hline
\endlastfoot
       0.0000000000000000   &    0.7071067811865475   &    0.7071067811865476\\ 
       0.0000000000000000   &    0.7071067811865476   &   -0.7071067811865475\\
       0.0000000000000000   &   -0.7071067811865475   &    0.7071067811865476\\
       0.0000000000000000   &   -0.7071067811865476   &   -0.7071067811865475\\
       0.7071067811865475   &    0.0000000000000000   &    0.7071067811865476\\
       0.7071067811865476   &    0.0000000000000000   &   -0.7071067811865475\\
      -0.7071067811865475   &    0.0000000000000001   &    0.7071067811865476\\
      -0.7071067811865476   &    0.0000000000000001   &   -0.7071067811865475\\
       0.7071067811865476   &    0.7071067811865475   &    0.0000000000000001\\
       0.7071067811865476   &   -0.7071067811865475   &    0.0000000000000001\\
      -0.7071067811865475   &    0.7071067811865476   &    0.0000000000000001\\
      -0.7071067811865475   &   -0.7071067811865476   &    0.0000000000000001\\
\end{longtable}
\end{center}

\begin{center}
\LTcapwidth=\textwidth
\begin{longtable}{rrr}
  \caption{Lebedev point subset 4/4.}
\label{tbl:LebSI4}
\\
\hline
       \multicolumn{3}{c}{set $B_1$} \\
  \hline
\endfirsthead
\multicolumn{3}{c}%
{\tablename\ \thetable\ -- \textit{Continued from previous page}} \\
  \hline
       \multicolumn{3}{c}{set $B_1$} \\
  \hline
\endhead
\hline \multicolumn{3}{r}{\textit{Continued on next page}} \\
\endfoot
\hline
\endlastfoot
       0.3015113445777636   &    0.3015113445777635   &    0.9045340337332909\\
       0.3015113445777637   &    0.3015113445777636   &   -0.9045340337332909\\
       0.3015113445777636   &   -0.3015113445777635   &    0.9045340337332909\\
       0.3015113445777637   &   -0.3015113445777636   &   -0.9045340337332909\\
      -0.3015113445777635   &    0.3015113445777636   &    0.9045340337332909\\
      -0.3015113445777636   &    0.3015113445777637   &   -0.9045340337332909\\
      -0.3015113445777635   &   -0.3015113445777636   &    0.9045340337332909\\
      -0.3015113445777636   &   -0.3015113445777637   &   -0.9045340337332909\\
       0.3015113445777636   &    0.9045340337332909   &    0.3015113445777636\\
       0.3015113445777636   &   -0.9045340337332909   &    0.3015113445777636\\
       0.3015113445777636   &    0.9045340337332909   &   -0.3015113445777635\\
       0.3015113445777636   &   -0.9045340337332909   &   -0.3015113445777635\\
      -0.3015113445777635   &    0.9045340337332910   &    0.3015113445777636\\
      -0.3015113445777635   &   -0.9045340337332910   &    0.3015113445777636\\
      -0.3015113445777635   &    0.9045340337332910   &   -0.3015113445777635\\
      -0.3015113445777635   &   -0.9045340337332910   &   -0.3015113445777635\\
       0.9045340337332909   &    0.3015113445777637   &    0.3015113445777636\\
      -0.9045340337332909   &    0.3015113445777637   &    0.3015113445777636\\
       0.9045340337332909   &    0.3015113445777637   &   -0.3015113445777635\\
      -0.9045340337332909   &    0.3015113445777637   &   -0.3015113445777635\\
       0.9045340337332909   &   -0.3015113445777637   &    0.3015113445777636\\
      -0.9045340337332909   &   -0.3015113445777637   &    0.3015113445777636\\
       0.9045340337332909   &   -0.3015113445777637   &   -0.3015113445777635\\
      -0.9045340337332909   &   -0.3015113445777637   &   -0.3015113445777635\\
\end{longtable}
\end{center}

\begin{center}
\LTcapwidth=\textwidth
\begin{longtable}{cc}
  \caption{Hydrogen cc-pVTZ/cc-pVTZ-RI species quadrature points (a.u.)}
\label{tbl:HptsSI}
\\
\hline
       H point set \\
  \hline
\endfirsthead
\multicolumn{1}{l}%
{\tablename\ \thetable\ -- \textit{Continued from previous page}} \\
  \hline
       H point set \\
  \hline
\endhead
\hline \multicolumn{1}{r}{\textit{Continued on next page}} \\
\endfoot
\hline
\endlastfoot
0.0 0.0 0.0 \\
$A_1$ $\otimes$ 0.2550495692028164\\
$A_2$ $\otimes$ 0.4414503914404356\\
$A_3$ $\otimes$ 0.5875929206946565\\
$A_1$ $\otimes$ 0.8500881858181466\\
$A_2$ $\otimes$ 0.8767919926845705\\
$A_3$ $\otimes$ 1.1376855741676248\\
$A_1$ $\otimes$ 1.4020561072906210\\
$A_2$ $\otimes$ 1.4031674307559165\\
$A_3$ $\otimes$ 1.6948362308564306\\
$B_1$ $\otimes$ 2.0508682846396238\\
$A_1$ $\otimes$ 2.3084240560165830\\
$A_2$ $\otimes$ 2.4853014001377569\\
$B_1$ $\otimes$ 2.7541840112271343\\
$A_3$ $\otimes$ 3.3784459365547317\\
$A_1$ $\otimes$ 3.9085612166141659\\
$A_2$ $\otimes$ 4.5534786191179117\\
\end{longtable}
\end{center}

\begin{center}
\LTcapwidth=\textwidth
\begin{longtable}{cc}
  \caption{Carbon cc-pVTZ/cc-pVTZ-RI species quadrature points (a.u.)}
\label{tbl:CptsSI}
\\
\hline
       C point set \\
  \hline
\endfirsthead
\multicolumn{1}{l}%
{\tablename\ \thetable\ -- \textit{Continued from previous page}} \\
  \hline
       C point set \\
  \hline
\endhead
\hline \multicolumn{1}{r}{\textit{Continued on next page}} \\
\endfoot
\hline
\endlastfoot
0.0 0.0 0.0 \\
$A_1$ $\otimes$ 0.0697148142120944\\
$A_2$ $\otimes$ 0.1453952389747425\\
$A_1$ $\otimes$ 0.2342968516739921\\
$A_2$ $\otimes$ 0.3167194257881547\\
$A_3$ $\otimes$ 0.3991658485572637\\
$A_1$ $\otimes$ 0.4921463978748848\\
$A_2$ $\otimes$ 0.5452940327070877\\
$A_3$ $\otimes$ 0.6928808595836493\\
$A_1$ $\otimes$ 0.7271808943305131\\
$A_2$ $\otimes$ 0.8364419359040032\\
$A_3$ $\otimes$ 0.9550764536689218\\
$B_1$ $\otimes$ 1.0042552897048800\\
$A_1$ $\otimes$ 1.2308082245100480\\
$A_2$ $\otimes$ 1.2515921948505666\\
$A_3$ $\otimes$ 1.3170945413732240\\
$B_1$ $\otimes$ 1.5668473153583786\\
$A_2$ $\otimes$ 1.7967421791285734\\
$A_1$ $\otimes$ 1.8562800941359989\\
$A_3$ $\otimes$ 1.9026574596444366\\
$B_1$ $\otimes$ 2.2023376401947057\\
$A_2$ $\otimes$ 2.4701298890775587\\
$A_1$ $\otimes$ 2.5826940890377630\\
$A_3$ $\otimes$ 2.6506301435069486\\
$B_1$ $\otimes$ 3.0810767533391212\\
$A_2$ $\otimes$ 3.4221828573728055\\
$A_3$ $\otimes$ 3.8904929263108845\\
$A_1$ $\otimes$ 4.0340344838211628\\
$A_2$ $\otimes$ 4.8577261225759178\\
$A_1$ $\otimes$ 5.2951816730851577\\
\end{longtable}
\end{center}

\begin{center}
\LTcapwidth=\textwidth
\begin{longtable}{cc}
  \caption{Nitrogen cc-pVTZ/cc-pVTZ-RI species quadrature points (a.u.)}
\label{tbl:NptsSI}
\\
\hline
       N point set \\
  \hline
\endfirsthead
\multicolumn{1}{l}%
{\tablename\ \thetable\ -- \textit{Continued from previous page}} \\
  \hline
       N point set \\
  \hline
\endhead
\hline \multicolumn{1}{r}{\textit{Continued on next page}} \\
\endfoot
\hline
\endlastfoot
0.0 0.0 0.0 \\
$A_1$ $\otimes$ 0.0740291727840731\\
$A_2$ $\otimes$ 0.1517711025394544\\
$A_1$ $\otimes$ 0.2292386711035654\\
$A_2$ $\otimes$ 0.2831000248738570\\
$A_3$ $\otimes$ 0.3644185443171212\\
$A_1$ $\otimes$ 0.4288518251583348\\
$A_2$ $\otimes$ 0.4799688137540949\\
$A_3$ $\otimes$ 0.5882996510447190\\
$A_1$ $\otimes$ 0.6142000220234174\\
$A_2$ $\otimes$ 0.6875219702081756\\
$B_1$ $\otimes$ 0.8246416514000185\\
$A_3$ $\otimes$ 0.8295086553115903\\
$A_1$ $\otimes$ 0.9989841825701087\\
$A_2$ $\otimes$ 1.0363846320748618\\
$A_3$ $\otimes$ 1.0745653323543152\\
$B_1$ $\otimes$ 1.2644981933147030\\
$A_2$ $\otimes$ 1.4818776161693972\\
$A_3$ $\otimes$ 1.5303532634022801\\
$A_1$ $\otimes$ 1.5656970159055199\\
$B_1$ $\otimes$ 1.7742342680710006\\
$A_2$ $\otimes$ 2.0608562398153891\\
$A_3$ $\otimes$ 2.0667401311291220\\
$A_1$ $\otimes$ 2.0980231282257580\\
$B_1$ $\otimes$ 2.4760916823114418\\
$A_2$ $\otimes$ 2.8170652651714376\\
$A_3$ $\otimes$ 3.1838378537539076\\
$A_1$ $\otimes$ 3.2554062082350321\\
$A_3$ $\otimes$ 4.0189700979204375\\
$A_1$ $\otimes$ 4.1136811865325891\\
\end{longtable}
\end{center}

\begin{center}
\LTcapwidth=\textwidth
\begin{longtable}{cc}
  \caption{Oxygen cc-pVTZ/cc-pVTZ-RI species quadrature points (a.u.)}
\label{tbl:OptsSI}
\\
\hline
       O point set \\
  \hline
\endfirsthead
\multicolumn{1}{l}%
{\tablename\ \thetable\ -- \textit{Continued from previous page}} \\
  \hline
       O point set \\
  \hline
\endhead
\hline \multicolumn{1}{r}{\textit{Continued on next page}} \\
\endfoot
\hline
\endlastfoot
0.0 0.0 0.0 \\
$A_1$ $\otimes$ 0.0652447257128732\\
$A_2$ $\otimes$ 0.1325503255253438\\
$A_1$ $\otimes$ 0.2050193077495113\\
$A_2$ $\otimes$ 0.2483307608247183\\
$A_3$ $\otimes$ 0.3296725279121236\\
$A_1$ $\otimes$ 0.3695681974012790\\
$A_2$ $\otimes$ 0.4317151572904991\\
$A_3$ $\otimes$ 0.5216903985417337\\
$A_1$ $\otimes$ 0.5314942941974161\\
$A_2$ $\otimes$ 0.6280697001504593\\
$B_1$ $\otimes$ 0.7138025771749562\\
$A_3$ $\otimes$ 0.8056864183075758\\
$A_1$ $\otimes$ 0.8746946212148640\\
$A_2$ $\otimes$ 0.9021904874234906\\
$A_3$ $\otimes$ 0.9801554595808909\\
$B_1$ $\otimes$ 1.0933052463892226\\
$A_2$ $\otimes$ 1.2908793119692796\\
$A_3$ $\otimes$ 1.3264294868555391\\
$A_1$ $\otimes$ 1.3499143280546431\\
$B_1$ $\otimes$ 1.5285596452003496\\
$A_2$ $\otimes$ 1.7616890999626422\\
$A_3$ $\otimes$ 1.8056343561817432\\
$A_1$ $\otimes$ 1.8079536046419200\\
$B_1$ $\otimes$ 2.1159638624942851\\
$A_2$ $\otimes$ 2.4301369998386630\\
$A_3$ $\otimes$ 2.6133556138762177\\
$A_1$ $\otimes$ 2.8080543419973161\\
$A_2$ $\otimes$ 3.3111645919590016\\
$A_1$ $\otimes$ 3.7734143044955140\\
\end{longtable}
\end{center}

\section{Octacene, decacene and $C_{60}$ geometries }
\label{secSI3}

\begin{center}
\LTcapwidth=\textwidth
\begin{longtable}{crrr@{\hskip 20mm}crrr}
  \caption{Octacene (B3LYP/6-31Gd geometry), Angstr\"{o}m}
\label{tbl:octaceneSI}
\\
\hline
\endfirsthead
\multicolumn{8}{c}%
{\tablename\ \thetable\ -- \textit{Continued from previous page}} \\
  \hline
\endhead
\hline \multicolumn{8}{r}{\textit{Continued on next page}} \\
\endfoot
\hline
\endlastfoot
C &    -3.714926 &    -1.409887 &     0.000000 & C &    -0.011113 &    -0.732348 &    -0.000001   \\
C &    -1.248764 &    -1.411259 &    -0.000000 & C &    -6.188589 &     1.412129 &     0.000000   \\
C &    -7.363538 &     0.717885 &     0.000001 & C &    -7.363537 &    -0.717890 &     0.000001   \\
C &    -6.188588 &    -1.412133 &     0.000001 & C &    -4.922719 &    -0.729554 &     0.000000   \\
C &    -4.922720 &     0.729551 &     0.000000 & C &    -3.714927 &     1.409885 &    -0.000000   \\
C &    -2.467867 &     0.731169 &    -0.000000 & C &    -2.467867 &    -0.731170 &    -0.000000   \\
C &    -1.248763 &     1.411258 &    -0.000000 & C &     6.142064 &    -1.411136 &    -0.000001   \\
C &     7.361111 &    -0.731209 &    -0.000000 & C &     7.361110 &     0.731206 &    -0.000000   \\
C &     6.142064 &     1.411132 &    -0.000000 & C &     4.904303 &     0.732349 &    -0.000001   \\
C &     4.904303 &    -0.732351 &    -0.000001 & C &     3.678182 &    -1.411821 &    -0.000001   \\
C &     2.446613 &    -0.732708 &    -0.000001 & C &     2.446614 &     0.732706 &    -0.000001   \\
C &     1.215008 &    -1.411813 &    -0.000001 & C &     3.678182 &     1.411817 &    -0.000001   \\
C &    -0.011112 &     0.732347 &    -0.000001 & C &     1.215010 &     1.411811 &    -0.000001   \\
H &     3.677950 &     2.500601 &    -0.000001 & H &     1.214600 &     2.500594 &    -0.000001   \\
H &     6.142287 &     2.499967 &    -0.000000 & H &    -3.715252 &    -2.498850 &     0.000000   \\
H &    -1.249306 &    -2.500074 &    -0.000000 & H &     6.142287 &    -2.499974 &    -0.000001   \\
H &     3.677949 &    -2.500606 &    -0.000001 & H &     1.214597 &    -2.500597 &    -0.000001   \\
H &    -6.186989 &     2.500307 &     0.000000 & H &    -8.312222 &     1.248416 &     0.000001   \\
H &    -3.715254 &     2.498849 &    -0.000000 & H &    -1.249305 &     2.500074 &    -0.000001   \\
H &    -6.186987 &    -2.500311 &     0.000001 & H &    -8.312221 &    -1.248423 &     0.000001   \\
C &    11.081835 &    -1.412118 &     0.000000 & C &    12.256783 &    -0.717907 &     0.000001   \\
C &    12.256783 &     0.717901 &     0.000001 & C &    11.081835 &     1.412113 &     0.000001   \\
C &     9.815989 &     0.729526 &     0.000000 & C &     9.815990 &    -0.729530 &     0.000000   \\
C &     8.608127 &    -1.409805 &    -0.000000 & C &     8.608127 &     1.409800 &     0.000000   \\
H &     8.607914 &     2.498780 &     0.000000 & H &    13.205553 &     1.248275 &     0.000001   \\
H &    11.079963 &     2.500292 &     0.000001 & H &    11.079964 &    -2.500297 &     0.000000   \\
H &    13.205553 &    -1.248280 &     0.000001 & H &     8.607914 &    -2.498787 &    -0.000000   \\
\end{longtable}
\end{center}

\begin{center}
\LTcapwidth=\textwidth
\begin{longtable}{crrr@{\hskip 20mm}crrr}
  \caption{Decacene (B3LYP/6-31Gd geometry), Angstr\"{o}m}
\label{tbl:decaceneSI}
\\
\hline
\endfirsthead
\multicolumn{8}{c}%
{\tablename\ \thetable\ -- \textit{Continued from previous page}} \\
  \hline
\endhead
\hline \multicolumn{8}{r}{\textit{Continued on next page}} \\
\endfoot
\hline
\endlastfoot
C &    -3.731097 &    -1.410141 &     0.000001 & C &    -0.026152 &    -0.732916 &    -0.000001   \\
C &    -1.265287 &    -1.411634 &    -0.000000 & C &    -6.204536 &     1.412316 &     0.000001   \\
C &    -7.379252 &     0.718055 &     0.000002 & C &    -7.379253 &    -0.718054 &     0.000003   \\
C &    -6.204537 &    -1.412316 &     0.000002 & C &    -4.938337 &    -0.729814 &     0.000001   \\
C &    -4.938337 &     0.729814 &     0.000001 & C &    -3.731097 &     1.410140 &     0.000000   \\
C &    -2.483272 &     0.731546 &    -0.000000 & C &    -2.483272 &    -0.731547 &     0.000000   \\
C &    -1.265286 &     1.411632 &    -0.000001 & C &     6.122035 &    -1.412588 &    -0.000002   \\
C &     7.349879 &    -0.733604 &    -0.000002 & C &     7.349879 &     0.733601 &    -0.000002   \\
C &     6.122034 &     1.412585 &    -0.000002 & C &     4.891019 &     0.733767 &    -0.000002   \\
C &     4.891020 &    -0.733772 &    -0.000002 & C &     3.660053 &    -1.412724 &    -0.000002   \\
C &     2.432210 &    -0.733597 &    -0.000002 & C &     2.432210 &     0.733590 &    -0.000002   \\
C &     1.197899 &    -1.412387 &    -0.000001 & C &     3.660052 &     1.412719 &    -0.000002   \\
C &    -0.026152 &     0.732911 &    -0.000001 & C &     1.197900 &     1.412382 &    -0.000001   \\
H &     3.659639 &     2.501473 &    -0.000002 & H &     1.197345 &     2.501150 &    -0.000001   \\
H &     6.122042 &     2.501360 &    -0.000002 & H &    -3.731487 &    -2.499098 &     0.000001   \\
H &    -1.265932 &    -2.500440 &    -0.000000 & H &     6.122044 &    -2.501362 &    -0.000002   \\
H &     3.659641 &    -2.501477 &    -0.000002 & H &     1.197343 &    -2.501153 &    -0.000001   \\
H &    -6.202983 &     2.500490 &     0.000001 & H &    -8.327996 &     1.248473 &     0.000003   \\
H &    -3.731487 &     2.499098 &     0.000000 & H &    -1.265929 &     2.500438 &    -0.000001   \\
H &    -6.202984 &    -2.500490 &     0.000002 & H &    -8.327996 &    -1.248472 &     0.000003   \\
C &     9.808174 &    -0.732855 &    -0.000001 & C &     8.584081 &    -1.412288 &    -0.000002   \\
C &     8.584080 &     1.412286 &    -0.000001 & C &    15.986667 &    -1.412324 &     0.000002   \\
C &    17.161322 &    -0.718096 &     0.000002 & C &    17.161321 &     0.718105 &     0.000003   \\
C &    15.986665 &     1.412331 &     0.000002 & C &    14.720357 &     0.729858 &     0.000001   \\
C &    14.720358 &    -0.729853 &     0.000001 & C &    13.513231 &    -1.410121 &     0.000000   \\
C &    12.265248 &    -0.731559 &    -0.000000 & C &    12.265248 &     0.731559 &    -0.000000   \\
C &    11.047375 &    -1.411596 &    -0.000001 & C &    13.513229 &     1.410124 &     0.000001   \\
C &     9.808174 &     0.732854 &    -0.000001 & C &    11.047374 &     1.411596 &    -0.000001   \\
H &    13.513335 &     2.499087 &     0.000001 & H &    11.047441 &     2.500404 &    -0.000000   \\
H &    18.110119 &     1.248420 &     0.000003 & H &    15.984999 &     2.500508 &     0.000002   \\
H &     8.583708 &    -2.501073 &    -0.000002 & H &    15.985003 &    -2.500500 &     0.000001   \\
H &    18.110122 &    -1.248409 &     0.000003 & H &    13.513339 &    -2.499083 &     0.000000   \\
H &    11.047442 &    -2.500403 &    -0.000001 & H &     8.583706 &     2.501072 &    -0.000001   \\
\end{longtable}
\end{center}

\begin{center}
\LTcapwidth=\textwidth
\begin{longtable}{crrr@{\hskip 20mm}crrr}
  \caption{C60 (B3LYP/6-311Gd geometry), Angstr\"{o}m}
\label{tbl:C60SI}
\\
\hline
\endfirsthead
\multicolumn{8}{c}%
{\tablename\ \thetable\ -- \textit{Continued from previous page}} \\
  \hline
\endhead
\hline \multicolumn{8}{r}{\textit{Continued on next page}} \\
\endfoot
\hline
\endlastfoot
C &     0.725748 &    -0.998906 &     3.321895 & C &    -0.725748 &    -0.998906 &     3.321895   \\
C &    -1.421859 &    -1.957021 &     2.589960 & C &    -0.696111 &    -2.955928 &     1.826864   \\
C &     0.696111 &    -2.955928 &     1.826864 & C &     1.421859 &    -1.957021 &     2.589960   \\
C &     1.174285 &     0.381548 &     3.321895 & C &     0.000000 &     1.234716 &     3.321895   \\
C &    -1.174285 &     0.381548 &     3.321895 & C &    -2.300617 &     0.747516 &     2.589960   \\
C &    -2.596144 &    -1.575473 &     1.826864 & C &    -1.421859 &    -3.191737 &     0.592148   \\
C &    -0.725748 &    -3.417918 &    -0.592148 & C &     0.725748 &    -3.417918 &    -0.592148   \\
C &     1.421859 &    -3.191737 &     0.592148 & C &     2.596144 &    -2.338570 &     0.592148   \\
C &     2.596144 &    -1.575473 &     1.826864 & C &     3.026364 &    -0.251390 &     1.826864   \\
C &     2.300617 &     0.747516 &     2.589960 & C &     0.000000 &     2.419011 &     2.589960   \\
C &     1.174285 &     2.800560 &     1.826864 & C &     2.300617 &     1.982232 &     1.826864   \\
C &     3.026364 &     1.746422 &     0.592148 & C &     3.474901 &     0.365967 &     0.592148   \\
C &     3.474901 &    -0.365967 &    -0.592148 & C &     3.026364 &    -1.746422 &    -0.592148   \\
C &     2.300617 &    -1.982232 &    -1.826864 & C &     1.174285 &    -2.800560 &    -1.826864   \\
C &    -2.596144 &    -2.338570 &     0.592148 & C &    -1.174285 &    -0.381548 &    -3.321895   \\
C &    -2.300617 &    -0.747516 &    -2.589960 & C &    -3.026364 &     0.251390 &    -1.826864   \\
C &    -2.596144 &     1.575473 &    -1.826864 & C &    -1.421859 &     1.957021 &    -2.589960   \\
C &     0.725748 &     0.998906 &    -3.321895 & C &     1.174285 &    -0.381548 &    -3.321895   \\
C &     0.000000 &    -1.234716 &    -3.321895 & C &     0.000000 &    -2.419011 &    -2.589960   \\
C &    -1.174285 &    -2.800560 &    -1.826864 & C &    -2.300617 &    -1.982232 &    -1.826864   \\
C &    -3.474901 &    -0.365967 &    -0.592148 & C &    -3.474901 &     0.365967 &     0.592148   \\
C &    -3.026364 &     1.746422 &     0.592148 & C &    -2.596144 &     2.338570 &    -0.592148   \\
C &    -1.421859 &     3.191737 &    -0.592148 & C &    -0.696111 &     2.955928 &    -1.826864   \\
C &     0.696111 &     2.955928 &    -1.826864 & C &     1.421859 &     1.957021 &    -2.589960   \\
C &     2.300617 &    -0.747516 &    -2.589960 & C &     3.026364 &     0.251390 &    -1.826864   \\
C &     2.596144 &     1.575473 &    -1.826864 & C &     2.596144 &     2.338570 &    -0.592148   \\
C &     1.421859 &     3.191737 &    -0.592148 & C &     0.725748 &     3.417918 &     0.592148   \\
C &    -0.725748 &     3.417918 &     0.592148 & C &    -1.174285 &     2.800560 &     1.826864   \\
C &    -2.300617 &     1.982232 &     1.826864 & C &    -3.026364 &    -1.746422 &    -0.592148   \\
C &    -3.026364 &    -0.251390 &     1.826864 & C &    -0.725748 &     0.998906 &    -3.321895   \\
\end{longtable}
\end{center}

\section{Relation with the LS-THC estimator}
\label{secSITHC}

The LS-THC estimator for expressing 2-electron Coulomb integrals using a real-space quadrature approach reads (eq.21 Ref.~\citenum{Parrish12}):

\begin{equation}
\argmin_{Z} \sum_{\rho,\rho'}\Big|\Big| (\rho|\rho') - \sum_{kk'} \rho(\mathbf{r}_k)\cdot Z_{kk'} \cdot \rho'(\mathbf{r}_{k'})\Big|\Big|^2,  \label{eq:fit_LSTHC}
\end{equation}

\noindent Adopting the RI-V approximation to express the $(\rho|\rho')$ Coulomb integrals:

\begin{equation}
\begin{split}
 (\rho|\rho')  & =  \sum_{\beta'} (\rho|\beta' )  \mathcal{F}_{\beta'}^{V} (\rho') \\ 
                          & =  \sum_{\beta'\beta'} \mathcal{F}_{\beta}^{V} (\rho)   (\beta|\beta' )  \mathcal{F}_{\beta'}^{V} (\rho') \\ 
\end{split}
\end{equation}

\noindent and using the previously introduced matrix notations $[D]_{k\rho}=\rho(r_k)$, $[F]_{\beta\rho}=\mathcal{F}_\beta^{V} (\rho)$ and writing $[V]_{\beta\beta'}= (\beta|\beta' )$, one obtains: 

\begin{equation}
[ (\rho|\rho')] = F^\dag \cdot V \cdot F
\end{equation}

\noindent and the  estimator in eq.\ref{eq:fit_LSTHC} reads:

\begin{equation}
\begin{split}
& \argmin_{Z} \Big|\Big|F^\dag \cdot V \cdot F - D^\dag \cdot  Z \cdot D \Big|\Big|^2 \\
= & \argmin_{Z} \, tr \Big[ (F^\dag \cdot V \cdot F - D^\dag \cdot  Z \cdot D)^\dag \cdot(F^\dag \cdot V \cdot F - D^\dag \cdot  Z \cdot D) \Big]\\
\label{eq::LS-THC_trace}
\end{split}
\end{equation}

\noindent Differentiating  expression \ref{eq::LS-THC_trace} with respect to $Z$  brings:

\begin{equation}
\begin{split}
& \frac{d}{dZ} \Big|\Big|F^\dag \cdot V \cdot F - D^\dag \cdot  Z \cdot D \Big|\Big|^2 \\
= & 2 \Big( D \cdot D^\dag \cdot Z \cdot D \cdot D^\dag - D \cdot F^\dag \cdot V \cdot F \cdot D^\dag ) =0
\end{split}
\end{equation}

\noindent resulting in the following expression for $Z$:

\begin{equation}
\begin{split}
Z & = ( D \cdot D^\dag)^{-1} \cdot D \cdot F^\dag \cdot V \cdot F \cdot D^\dag \cdot ( D \cdot D^\dag)^{-1}\\
& = M^\dag \cdot V \cdot M\\
\end{split}
\end{equation}

\noindent with $M$ the fitting matrix as defined by equation (8) of our main manuscript:

\[
M = F\cdot D^\dag \cdot ( D\cdot D^\dag )^{-1}\label{eq:fit_LSQR_3}
\]

\noindent We can thus see that the formulation of the equation (6) of our main manuscript allows to recover the LS-THC estimator of eq.\ref{eq:fit_LSTHC} with a $\mathcal{O}(N^3)$ computational effort.

\nocite{*}
\bibliography{supportingInfo}